\documentclass[iop,twocolappendix]{emulateapj}
\usepackage{epsfig}
\usepackage{mathptmx} 
\usepackage{graphicx}
\usepackage[breaklinks=true]{hyperref} 

\shorttitle{Dwarf Galaxies with Radiation Feedback. II}
\shortauthors{J. Kim et al.}

\begin{document}

\title{Dwarf Galaxies with Ionizing Radiation Feedback. II: Spatially-resolved Star Formation Relation}

\author{Ji-hoon Kim \altaffilmark{1,2}}
\email{me@jihoonkim.org}
\author{Mark R. Krumholz \altaffilmark{1}}
\author{John H. Wise \altaffilmark{3}} 
\author{Matthew J. Turk \altaffilmark{4}} 
\author{Nathan J. Goldbaum \altaffilmark{1}}
\author{Tom Abel \altaffilmark{2}}

\altaffiltext{1}{Department of Astronomy and Astrophysics, University of California, Santa Cruz, CA, USA}
\altaffiltext{2}{Kavli Institute for Particle Astrophysics and Cosmology, Stanford University, Stanford, CA, USA}
\altaffiltext{3}{Center for Relativistic Astrophysics, School of Physics, Georgia Institute of Technology, Atlanta, GA, USA}
\altaffiltext{4}{Department of Astronomy and Astrophysics, Columbia University, New York, NY, USA}

\begin{abstract}
We investigate the spatially-resolved star formation relation using a galactic disk formed in a comprehensive high-resolution (3.8 pc) simulation.
Our new implementation of stellar feedback includes ionizing radiation as well as supernova explosions, and we handle ionizing radiation by solving the radiative transfer equation rather than by a subgrid model.
Photoheating by stellar radiation stabilizes gas against Jeans fragmentation, reducing the star formation rate. 
Because we have self-consistently calculated the location of ionized gas, we are able to make simulated, spatially-resolved observations of star formation tracers, such as ${\rm H}\alpha$ emission.  
We can also observe how stellar feedback manifests itself in the correlation between ionized and molecular gas.
Applying our techniques to the disk in a galactic halo of $2.3 \times 10^{11} M_{\odot}$, we find that the correlation between star formation rate density (estimated from mock ${\rm H}\alpha$ emission) and ${\rm H}_2$ density shows large scatter, especially at high resolutions of $\lesssim$ 75 pc that are comparable to the size of giant molecular clouds (GMCs).
This is because an aperture of GMC size captures only particular stages of GMC evolution, and because ${\rm H}\alpha$ traces hot gas around star-forming regions and is displaced from the ${\rm H}_2$ peaks themselves.  
By examining the evolving environment around star clusters, we speculate that the breakdown of the traditional star formation laws of the Kennicutt-Schmidt type at small scales is further aided by a combination of stars drifting from their birthplaces, and molecular clouds being dispersed via stellar feedback.
\end{abstract}


\section{Introduction}\label{sec-IV:1}

Understanding how galaxies have evolved into their current shapes is one of the most intensely pursued, yet one of the most poorly understood topics of contemporary astrophysics.   
By definition the process of galaxy formation cannot be separated from the problem of star formation.  
Any analytic, semi-analytic, or numerical framework to build galaxies requires a model to form stars and describe their feedback. 
But, how good is our understanding of star cluster physics, especially when the star formation and feedback themselves cannot be understood outside the galactic context? 

One may attempt to verify the star formation models by employing them in numerical studies of galaxy formation, and then comparing the resulting galaxies with the gold standard we already possess: observed galaxies.  
As a matter of fact, we already have an abundant supply of galactic observations accumulated over decades to which we can compare our simulated galaxies, including photometric images of galaxies at various wavelengths, spectral energy distributions, star formation histories, and star formation relations.
In particular, the well-known Kennicutt-Schmidt star formation relation connects the surface densities of gas and star formation rate \citep{1959ApJ...129..243S, 1989ApJ...344..685K, 1998ApJ...498..541K, 2007ApJ...671..333K}. 
Mounting observational evidence and theoretical investigations now strongly suggest that the star formation rate is also, or even more closely, related with molecular gas density \citep[e.g.][]{2002ApJ...569..157W, 2007ApJ...669..289K, 2008ApJ...680.1083R, 2008AJ....136.2846B, 2009ApJ...699..850K, 2010AJ....140.1194B, 2011ApJ...730L..13B, 2011ApJ...741...12B, 2011AJ....142...37S, 2011ApJ...732..115F, 2012ApJ...745...69K, 2013arXiv1301.2328L}, which will be the focus of the study presented in this paper.  

Yet, in order to make a reliable and meaningful comparison between simulations and observations for the purpose of understanding galactic star formation, one must ensure that the galaxy simulation be properly coupled with star cluster physics.  
To that end, the following conditions should be met: 
{\it (a)} {\it Sufficient resolution in time and space} to describe the relevant star cluster physics.  
Resolution of $\lesssim$ 10 pc is needed to resolve the scale of star-forming molecular clouds and to produce a multiphase interstellar medium (ISM).
A corresponding timestep would resolve the lifetime of massive stars and molecular clouds \citep[e.g.][]{2010ApJ...720L.149T, 2010MNRAS.409.1088B}.  
{\it (b)} {\it Physically-motivated models for star cluster formation and feedback} with as little subgrid physics as possible. 
Models should correspond with the spatial and temporal resolution of the simulation. 
For example, a star formation model that solely depends upon the star formation relation observed at $\sim$ kpc scales \citep[e.g. $\Sigma_{\rm SFR} \sim \Sigma_{\rm gas}^{1.4}$;][]{1998ApJ...498..541K} may not be desired in high-resolution simulations.
Similarly, a phenomenological feedback model such as stopping the cooling near a star cluster particle right after its birth might not be a suitable choice given fine temporal resolution \citep[see, however,][]{2011MNRAS.417..950H, 2012arXiv1208.0002S}.  
Not surprisingly, simulations at this level of sophistication require enormous computational costs.  
For this reason one may arguably state that no simulation has reliably checked the physics of galactic star formation against observed galaxies.  
Indeed, galactic star formation still remains to be fully understood.  

\begin{table*}[t]  
\caption[Simulation suite description]{Simulation Suite Description}
\centering     
\begin{tabular}{l  ||  c c c } 
\hline\hline   
\multicolumn{1}{c ||}{Physics\tablenotemark{a}} & MC-TF  & MC-RTF  \\ [1ex] 
\hline      
Star-forming molecular could (SFMC) particle formation        &\textcircled{}&\textcircled{}\\    
SFMC feedback: supernova explosion        &\textcircled{}&\textcircled{}\\    
SFMC feedback: ionizing radiation              &$\times$&\textcircled{}\\  [1ex] 
\hline
\end{tabular} 
\tablenotetext{1}{\scriptsize For detailed explanation, see \S\ref{sec-IV:2} or Paper I.  $\circ$ = included, $\times$ = not included.}
\label{table-IV:desc}  
\end{table*}

Thanks to the expeditious developments of both numerical techniques and computer hardware, however, simulations with significantly improved resolution are now feasible at increasingly cheaper costs. 
Utilizing adaptive mesh refinement techniques, in the first paper of this series \citep[][ hereafter Paper I]{2013ApJ...775..109K} we simulated the behavior of star clusters {\it and} their embedding galaxy in a single self-consistent numerical framework.
The implementation includes a new way of describing feedback from star-forming molecular cloud (SFMC) particles by combining ionizing stellar radiation and supernova explosions.  
The radiation feedback from each of $\lesssim$10000 SFMC particles is rendered by tracing the ultraviolet photon rays on the fly.  
This means that our approach does not rely on a subgrid model to handle the stellar radiation feedback, but actually solves the transfer equation.  
Joined with 3.8 pc resolution, this new scheme allowed the authors to study the escape of ionizing photons from star-forming clumps and from a galaxy.  
By simulating a galactic halo of $2.3\times10^{11} M_{\odot}$, we found that the galactic escape fraction is dominated by a small number of molecular clouds with exceptionally high escape fraction. 
We also discovered that the escape fraction from a SFMC particle rises on average from 0.27\% at its birth to 2.1\% at the end of a cloud lifetime, 6 Myrs.  

The machinery established in Paper I provides us with a unique opportunity to look into the physics of star formation and its intertwined nature with the galactic context.  
Moreover, our stellar radiation feedback model enables us to make unprecedented simulated observations of star formation tracers, such as ${\rm H}\alpha$ emission.
In the second paper of the series, we use the numerical framework developed in Paper I to carry out an analysis of self-regulated star formation in a dwarf-sized galaxy.  
We make a direct comparison of star formation tracers between simulated galaxies and observed ones.  
Special attention is paid to the spatially-resolved star formation relation observed in local galaxies, and why it breaks down at small scales.  
We demonstrate how galactic star formation can be better understood by significantly expanding the usage of cutting-edge numerical simulations.    

This article is organized as follows.  
The physics in the simulation code and the initial condition are briefly explained in \S \ref{sec-IV:2} and \S \ref{sec-IV:3}, respectively.  
\S \ref{sec-IV:4} overviews the global star formation of the performed runs. 
\S \ref{sec-IV:5} is devoted to the spatially-resolved star formation relation via simulated maps of physical observables such as ionized gas tracers, ${\rm H}{\alpha}$.  
We also inspect the evolving environment of star-forming gas clumps.
Assembled in \S \ref{sec-IV:6} are the summary and conclusions.  

\begin{figure*}[t]
\epsscale{1.15}
\plotone{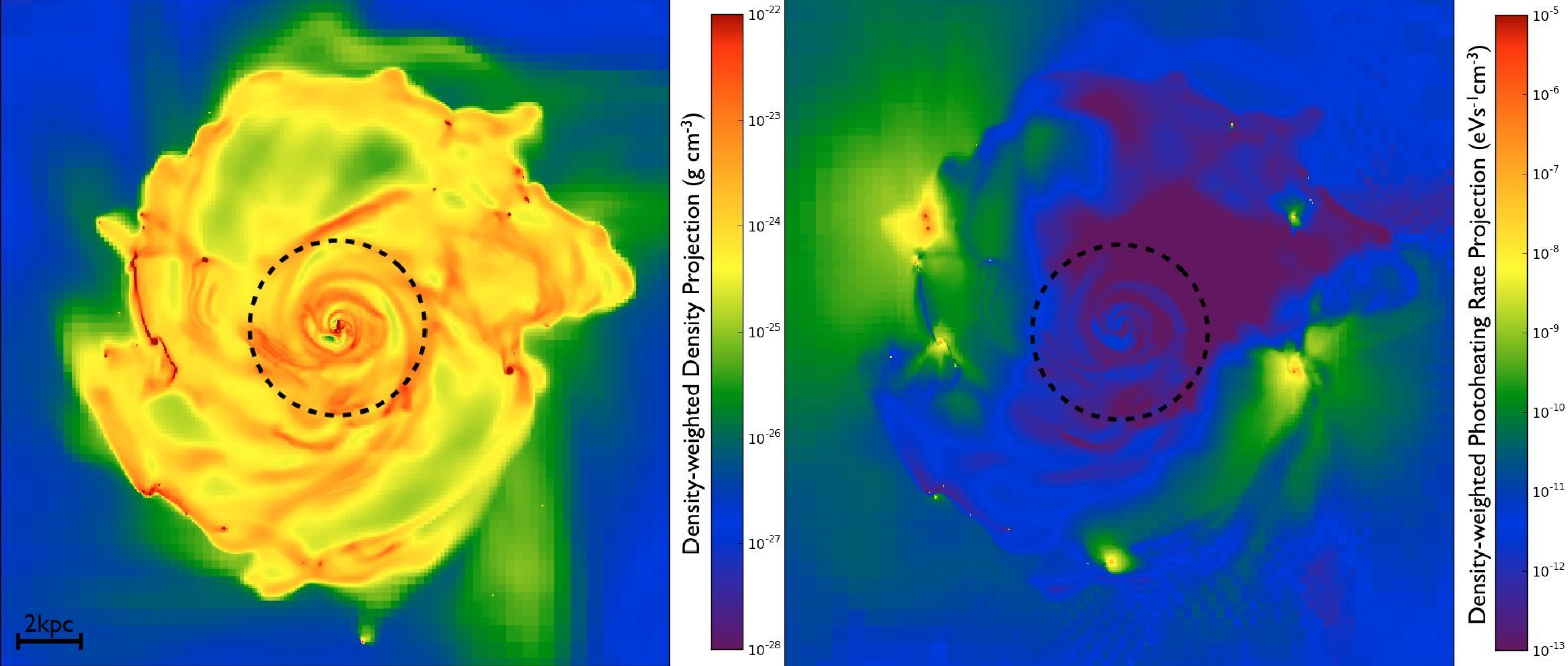}
    \caption{{\it Left:} face-on density-weighted projection of density in the MC-RTF run in a 20 kpc box at 13.3 Myr into the high-resolution evolution.  High-density clumps which bear a number of SFMC particles are noticeable.  
 {\it Right:} density-weighted projection of photoheating rates.  Note that the photoheating rates are low in the galactic center ($<$ 2.5 kpc in radius, dotted circles) because the star-forming molecular cloud (SFMC) particles there are not ray-emitting.   
\label{fig:gammaHI_eVscm3_up}}
\end{figure*}

\section{Physics In The Code}\label{sec-IV:2}

Our enhanced version of the Eulerian adaptive mesh refinement  {\it Enzo-2.1}\,\footnote{http://enzo-project.org/} \citep{1997ASPC..123..363B, 2007arXiv0705.1556N, 2011MNRAS.414.3458W, 2009ApJ...694L.123K, 2011ApJ...738...54K} contains relevant physics previously considered in galactic simulations as well as the new treatments of SFMC particle feedback discussed in detail in Paper I.
The framework developed in Paper I is designed to investigate galactic star formation of a dwarf-sized galaxy in a numerical simulation, without omitting the detailed interaction between the SFMC particles and their surrounding galactic context. 
While the key ingredients of the framework are detailed in Paper I, we briefly review its major features focusing on the SFMC feedback model. 
Table \ref{table-IV:desc} also summarizes the feedback models we have employed in the reported study.  

The ZEUS hydrodynamics module in {\it Enzo} is employed to solve the Euler equations for collisional fluid \citep{1992ApJS...80..753S, 1992ApJS...80..791S}.
We use {\it Enzo}'s multi-species non-equilibrium chemistry and cooling modules to track reactions among six species (H, ${\rm H}^+$, He, ${\rm He}^+$, ${\rm He}^{++}$, ${\rm e}^-$) and radiative losses. 
Cooling by metals is included in gas above $10^4$ K using the cooling rates tabulated by \cite{1993ApJS...88..253S}, and in gas below $10^4$ K using the cooling rate approximation of \cite{2002ApJ...564L..97K}.
The mass thresholds above which a cell is refined by factors of 2 in each axis are functions of a refinement level {\it l} as $M_{\rm ref, gas}^{l} = 2^{-0.820 l} \cdot 4.87\times 10^6 \,M_{\odot}$ for gas, and $M_{\rm ref, part}^{l} = 2^{-0.533 l} \cdot 4.87\times 10^6 \,M_{\odot}$ for particles, respectively.  
At the finest level $l=13$ of $\Delta\, x = 3.8 \,\,{\rm pc}$ resolution, a cell would have been refined if more than $3000 \,\,M_{\odot} $ in gas or $40000 \,\,M_{\odot} \simeq 2.5 \,\, M_{\rm DM}$ in particles.  
This resolution is in accord with the Jeans length for a dense gas clump of $n=1000$ ${\rm cm^{-3}}$ at $\sim$ 100 K, at which point a corresponding Jeans mass of $2000 \,\,M_{\odot}$ collapses to spawn a SFMC particle.
Indeed a finest cell produces a SFMC particle of initial mass $M_{\rm MC}^{\rm init} = 0.5 \rho_{\rm gas} \Delta \,x^3$ when {\it (a)} the proton number density exceeds the threshold $1000$ ${\rm cm^{-3}}$, {\it (b)} the velocity flow is converging, {\it (c)} the cooling time $t_{\rm cool}$ is shorter than the gas dynamical time $t_{\rm dyn}$ of the cell, and {\it (d)} the particle produced has at least $M_{\rm thres}  = 1000 \,\,M_{\odot}$.
This formulation guarantees that a SFMC particle is created before an unresolved gas clump interferes with the consistency of hydrodynamics.  

In our simulation, the later evolution of a SFMC particle describes three key aspects of star formation in Giant Molecular Clouds (GMCs):
{\it (a)} $0.24\,\,M_{\rm MC}^{\rm init}$ is instantly turned into stars, but the rest is considered as atomic or molecular gas that does not participate in star formation.  
{\it (b)} Among stars, $0.04\,\,M_{\rm MC}^{\rm init}$ is massive enough to commence Type II supernova peaking at the particle age of 5 Myr.  
The rest of the stellar mass, $0.2\,\,M_{\rm MC}^{\rm init}$, is considered as {\it long-lived} stellar mass, $M_{*}$. 
{\it (c)} $0.76+0.04 = 0.8\,\,M_{\rm MC}^{\rm init}$ is eventually recycled back into the ISM with the energy of exploding supernovae ($7.5 \times 10^{-7}$ of the rest mass energy of $0.2\,M_{\rm MC}^{\rm init}$).   
2\% of the ejected mass is considered as metals enriching the ISM. 
Overall, the mass of a SFMC particle evolves as 
\begin{eqnarray}
M_{\rm MC}(t) = M_{\rm MC}^{\rm init}   \left(1 - 0.8 \left[0.5 - 0.5\,\,{\rm erf} \{ \sqrt{8}(5 - T) \}  \right] \right),
\label{eq:SFMC_mass}
\end{eqnarray}
where ${\tt erf}$() is the Gauss error function, and $T = t-t_{\rm cr}$ is the particle age in Myr with particle creation time $t_{\rm cr}$.  

In order to describe the ionizing ultraviolet photons from young star clusters, we perform a self-consistent transport calculation of monochromatic stellar radiation from SFMC particles \citep{2002MNRAS.330L..53A, 2011MNRAS.414.3458W, 2011ApJ...738...54K}.  
From $T = 0$ to 6 Myrs after the birth, ionizing radiation luminosities $L_{\rm MC}(i,t) = q_{\rm MC}  \, E_{\rm ph}  \, M_{\rm MC}(i, t)$ are assigned to the SFMC particles that are more than 2.5 kpc away from the galactic center. 
Here $q_{\rm MC} = 6.3 \times 10^{46} \,\,{\rm photons} \,\, {\rm s}^{-1} M^{-1}_{\odot}$ is the lifetime-averaged ionizing luminosity per solar mass in clusters \citep{2010ApJ...709..424M}, and $E_{\rm ph} = 16.0\, {\rm eV}$ is the mean energy per deposited photon \citep{2006ApJS..162..281W}.\footnote{By using the particle mass, $M_{\rm MC}(i,t)$, rather than the stellar mass inside we aim to approximately encapsulate various other types of feedback beyond photoionization, such as protostellar outflows, stellar winds, and radiation pressure.  More discussion on this in Paper I.}  
The SFMC particles in the inner 2.5 kpc disk do not emit radiation so that we concentrate on the evolution of clouds in the galactic spiral arms and outer disk, and expedite the radiative transfer calculation.
Each of the $12\times4^3$ rays isotropically cast from the particle is traced until most of its photons are absorbed or until it reaches the edge of the computational domain, while being adaptively split into child rays whenever the angular resolution associated with it grows larger than $0.2\, (\Delta x)^2$ of a local cell.\footnote{Other techniques to speed up the radiation calculation include:  {\it (a)} A ray is no longer split when it is more than 5 kpc away from the source, a distance far enough for a dwarf-sized galaxy.  {\it (b)} Two rays merge to form a single ray when the distance from the source is more than 10 times the separation between the two sources.  {\it (c)} Two radiating SFMC particles merge if they are separated by less than 5 pc in space, and $3 \times 10^4$ yr in age.}
Photons in the emitted ray interact with hydrogen in the surrounding gas in three different ways: they (1) {\it photoionize} hydrogen (2) {\it photoheat} the gas with the excess energy above the ionization threshold, and (3) exert outward {\it momentum} when they are absorbed by the gas cell. 
Dust absorption, or the infrared photons re-emitted by UV-irradiated dust grains and the momentum imparted by them are not explicitly included in our calculation. 

\section{Initial Conditions}\label{sec-IV:3}

We construct a dark matter halo of $2.3\times10^{11} M_{\odot}$ that follows the Navarro-Frenk-White profile with a concentration $c = 10$ and a spin parameter $\lambda = 0.05$.  
Gas grids are generated by splitting particles with an initial baryonic mass fraction of 10\% and metallicity of $Z_{\rm init} = 0.003\,\,Z_{\odot}$.  
We then produce a galaxy embedded in a dark matter halo by performing a coarse resolution (15.2 pc) simulation for $\sim$ 1 Gyr in a $32^3$ root grid box of $1\,{\rm Mpc}^3$. 
When a relaxed, well-defined galactic disk has emerged, $\sim 1.3\times10^{10} M_{\odot}$ is in $3.4\times10^{6}$ star cluster particles formed at low resolution, and $\sim 6.5\times10^{9} M_{\odot}$ is in gaseous form, all embedded in a halo of $1.3\times 10^7$ dark matter particles ($16000\,\, M_{\odot}$ each). 
We now employ high-resolution refinement criteria (3.8 pc, 13 additional levels of refinement) to resolve the galaxy down to the size of a molecular cloud, and make the entire galaxy evolve for $\sim$ 30 Myrs.  
As a consequence, the computational expense of performing a high resolution calculation for a galactic dynamical time is saved, but one may still observe the star-forming gas clumps for their typical lifetime.
To exclude the period in which the galaxy is trying to reach new equilibrium with freshly imposed high resolution, we investigate the galaxy from 10 to 30 Myr into the evolution in the remainder of this article.
Figure \ref{fig:gammaHI_eVscm3_up} shows the snapshot of our simulated galaxy in the MC-RTF run at 13.3 Myr into the high-resolution evolution.

\section{Results I: Global Star Formation} \label{sec-IV:4}

We analyze two simulations listed in Table \ref{table-IV:desc} to examine the global star formation, and the spatially-resolved star formation relation in a simulated dwarf-sized galaxy.
First we briefly overview the global star formation rate on the galactic disk and the structure of the galactic ISM. 

\subsection{Self-regulated Galactic Star Formation}  \label{sec-IV:4-SF}

\begin{figure}[t]
\epsscale{1.14}
\plotone{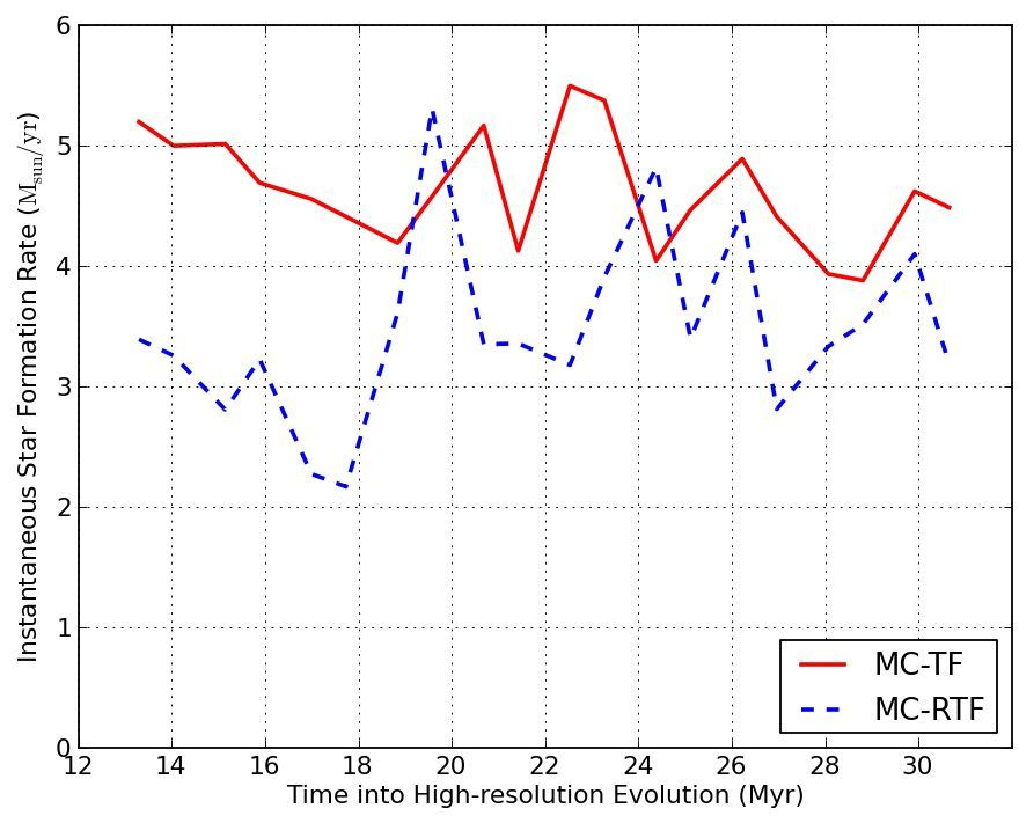}
    \caption{Time evolution of the instantaneous galactic star formation rate in the unit of $M_{\odot} {\rm yr}^{-1}$ from 13.3 Myr to 30.7 Myr of the high-resolution evolution.  For the MC-TF ({\it solid line}, only with supernova feedback) and MC-RTF runs ({\it dashed line}, with both supernova and stellar radiation feedback).  
\label{fig:sfr_temporal}}
\end{figure}

Shown in Figure \ref{fig:sfr_temporal} is the instantaneous galactic star formation rate, $dM_*/dt$ (with $dt = 0.37$ Myr), of the run with only supernova feedback, MC-TF, and the run with both radiation and supernova feedback, MC-RTF.
During the $\sim$ 20 Myrs evolution with 3.8 pc resolution that is analyzed, the star formation rate (SFR) on average is $4.60\,\,M_{\odot}{\rm yr}^{-1}$ and $3.57\,\,M_{\odot}{\rm yr}^{-1}$, respectively. 
In other words, the MC-RTF run forms 22.4\% less stellar mass than the MC-TF run does.
Our realistic description of stellar feedback helps to self-regulate galactic star formation, and achieve reasonable star formation rates during the analyzed period.  

\begin{figure}[t]
\epsscale{1.19}
\plotone{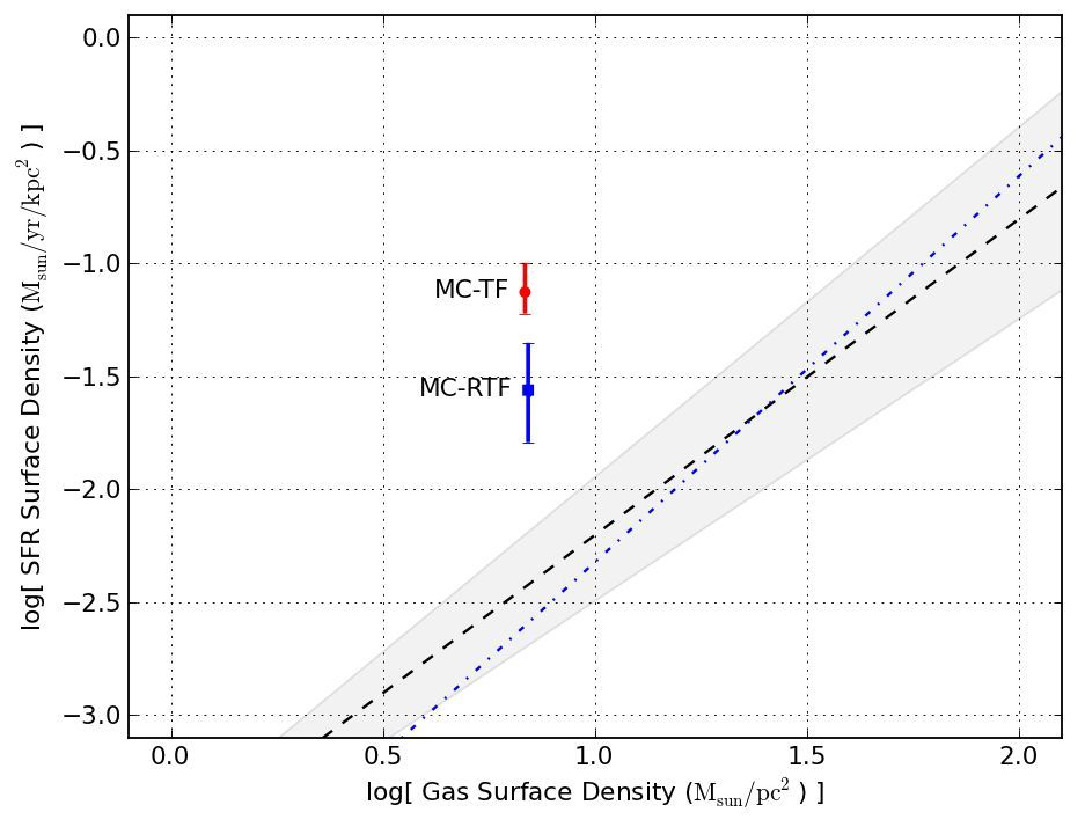}
    \caption{Time-averaged location of global SFR surface density, $\overline \Sigma_{\rm SFR}$, and global gas surface density, $\overline \Sigma_{\rm gas}$, between 13.3 Myr and 30.7 Myr of the high-resolution evolution, for the MC-TF ({\it red circle}) and MC-RTF runs ({\it blue square}). Also plotted are the observed global Kennicutt-Schmidt relations: the best fits to the disk-averaged galactic observations by \citet[][{\it long dashed line} with a shaded region representing their errors]{1998ApJ...498..541K} and by \citet[][{\it dot dashed line} that includes high-$z$ galaxies]{2007ApJ...671..303B}.  
\label{fig:global_ks}}
\end{figure}

Figure \ref{fig:global_ks} illustrates the time-averaged location of the global SFR surface density, $\overline \Sigma_{\rm SFR}$, and global gas surface density, $\overline \Sigma_{\rm gas}$, for the two runs with and without the stellar radiation feedback, between 13.3 Myr and 30.7 Myr of the high-resolution evolution.
Each of the densities is measured on a galactic disk of 9 kpc in radius excluding the central 2.5 kpc in which SFMC particles do not radiate ionizing photons (see \S\ref{sec-IV:2}). 
We estimate the SFR surface density by using the stars of age less than $T = 0.37$ Myr.
The movement of the data points in time on this plane is relatively small during the analyzed period which covers only 17.4 Myrs.  
The error bars here show the extent of such movements; and the movement in $x$-axis (variation in $\overline \Sigma_{\rm gas}$) is so small that its error bar is not visible in the plot.  
We recognize that the additional suppression by stellar radiation is not large enough to move the data points down to the observed global Kennicutt-Schmidt relation \citep{1998ApJ...498..541K}.  
The global star formation surface density would have been lower if a larger energy input had been employed for thermal supernovae explosion (see \S\ref{sec-IV:2} and/or Paper I; we intentionally choose a relatively small fiducial value for thermal feedback in order to further contrast the effect of radiation feedback). 
The suppression by feedback could have been even more effective if the SFMC particles in the inner disk had also radiated ionizing photons ($<$ 2.5 kpc from the galactic center).

\subsection{Interstellar Medium Structure}  \label{sec-IV:4-ISM}

\begin{figure*}[t]
\epsscale{1.18}
\plotone{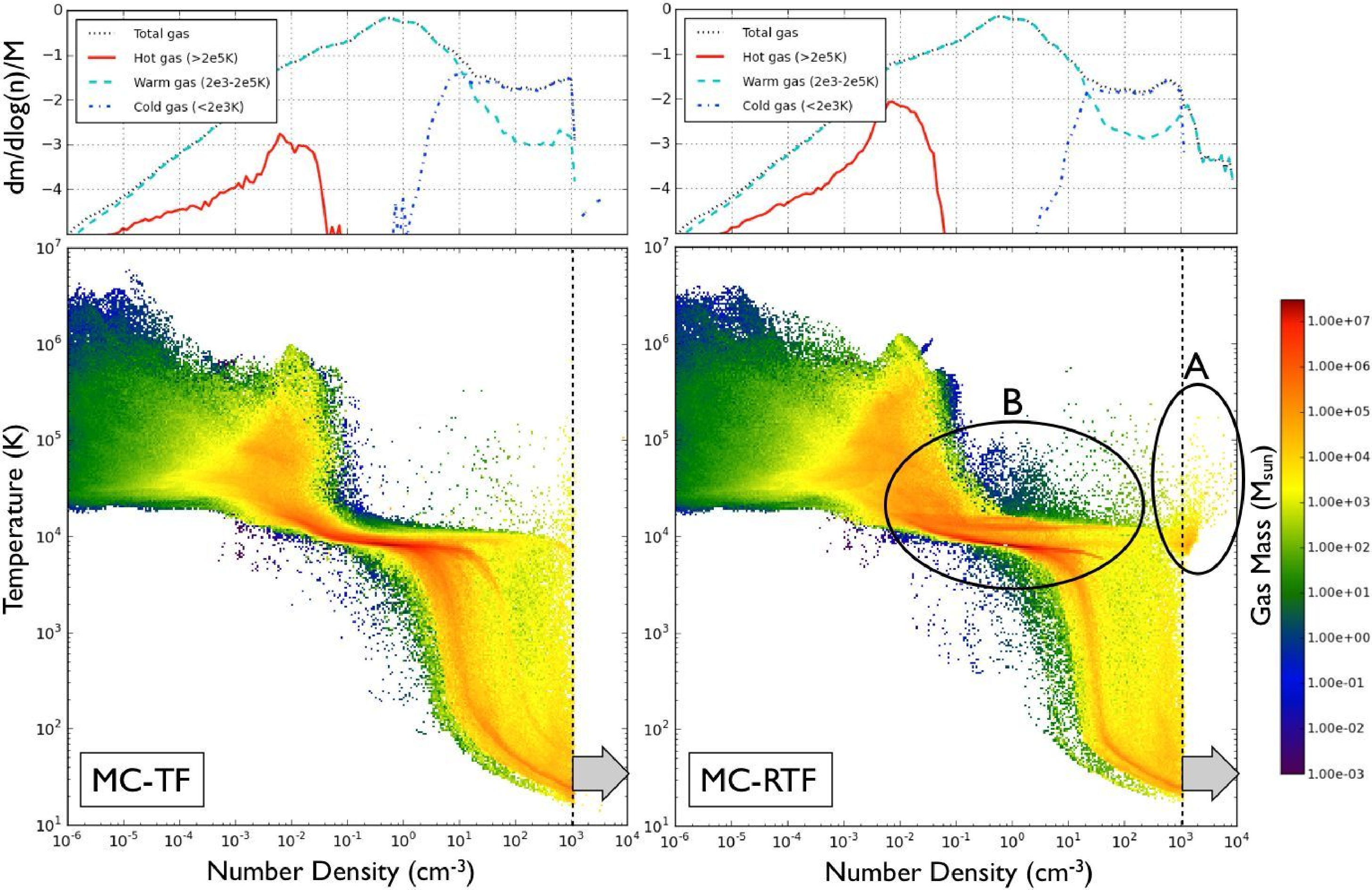}
    \caption{{\it Bottom rows:} joint probability distribution functions (PDFs) of gas number density and temperature colored by mass in each bin, for a 10 kpc sphere from a galactic center (but excluding the inner 2.5 kpc sphere) at 13.3 Myrs into the high-resolution evolution ({\it left:} MC-TF run, only with supernova feedback, {\it right:} MC-RTF run, with supernova and radiation feedback).  The vertical dashed line in each plot denotes the density threshold for SFMC particle formation.  The ionizing photons from SFMC particles heat the surrounding dense gas up to $\sim 10^5$ K and prevent star formation.  It increases the amount of very dense gas  which otherwise could have been gravitationally unstable  ($> 10^3 \,\,{\rm cm}^{-3}$; zone ``A'').   {\it Top rows:} one dimensional PDFs of gas number density that share the $x$-axes with the bottom rows.  In the MC-RTF run, a significant amount of warm gas ($2\times10^3 - 2\times10^5\,\,{\rm K}$) is existent above $10^3 \,\,{\rm cm}^{-3}$.  
\label{fig:PDF_compare}}
\end{figure*}

We now focus on how ionizing radiation feedback regulates star formation.  
Photoionizing radiation from massive stars is one of the important factors that drives changes in the environment of star-forming clumps, and self-regulates star formation \citep{1979MNRAS.186...59W, 2002ApJ...566..302M}.
In order to characterize the nature of the impact of ionizing stellar radiation feedback, we compare the galactic ISM structures of the two runs, MC-TF and MC-RTF.  
The bottom rows of Figure \ref{fig:PDF_compare} display the two dimensional probability distribution functions (PDFs) of proton number density and gas temperature, colored by gas mass in each bin.  
The measurement is made in the sphere of 10 kpc radius centered on a galactic center (but excluding the inner 2.5 kpc sphere) at 13.3 Myrs into the high-resolution evolution for each of the performed simulations.  

The ionizing radiation from young SFMC particles heats the surrounding dense gas up to $\sim 10^5$ K and pushes the PDF in an upward direction.  
In the MC-TF run dense gas cells with number density above $10^3 \,\,{\rm cm}^{-3}$, the threshold density for the SFMC particle formation  (see \S \ref{sec-IV:2}), are almost completely turned into SFMC particles.  
However, in the MC-RTF run, zone ``A'' still hosts very dense gas cells above $10^3 \,\,{\rm cm}^{-3}$, which are beyond the threshold density.  
Photoheating by stellar radiation retains such hot dense gas which otherwise would have been unstable against Jeans fragmentation and deposited into SFMC particles.  
At the same time, zone ``B'' depicts relatively less dense gas cells heated up to $\sim 10^5$ K, but mostly clustered around $\sim2 \times 10^4$ K just above the bump in the cooling curve by the [OIII] forbidden line.  
The top rows of the same figure show one dimensional PDFs of gas number density at the same epoch, binned by cold ($< 2\times10^3 \,\,{\rm K}$), warm ($2\times10^3 - 2\times10^5\,\,{\rm K}$), and hot gas ($>2\times10^5\,\,{\rm K}$). 
In the plot for the MC-RTF run, the increased masses of the warm gas above $10^3 \,\,{\rm cm}^{-3}$ (corresponding to zone ``A'') and the hot gas at around $\sim 10^{-2} \,\,{\rm cm}^{-3}$ are prominent.

\section{Results II: Spatially-resolved Star Formation Relation} \label{sec-IV:5}

In this section we move to the spatially-resolved star formation relation on a simulated galactic disk.  
We start by describing our methodology which allows us to make spatially-resolved mock observations of galactic star formation tracers, such as ${\rm H}\alpha$ emission. 

\subsection{Simulated Observation of Star Formation Relation}  \label{sec-IV:5-obs}

The traditional star formation tracer of ${\rm H}\alpha$ emission (Balmer line of $n=3  \rightarrow 2, \,\,\,1.88 \,{\rm eV}$) maps out the ionized gas around young massive stars.  
The site of highly ionized gas, however, may not overlap with the site of dense collapsing gas in which on-going star formation takes place, usually identified by peaks of molecular hydrogen, ${\rm H}_2$.  
This discrepancy may become even more problematic when the observed galaxy is resolved with higher spatial accuracy which distinguishes such regions from one another.  
Therefore in an observation with high spatial resolution, one may question the validity of the traditional star formation tracers, in particular, ${\rm H}\alpha$ emission \citep[e.g.][]{2010ApJ...722L.127O, 2010ApJ...722.1699S,2012ApJ...757..138K}.
To understand and characterize this phenomenon, we now produce {\it mock observations} of various physical properties in our simulated galaxy. 
Our galaxy is an ideal laboratory for such an exercise not only because we have self-consistently tracked radiation fields like photoionization rates prompted by $\lesssim$10000 SFMC particles with non-equilibrium primordial chemistry, but because the galaxy is simulated with high resolution, 3.8 pc, better than any reported observation of local galaxies.  
A high dynamic range in our simulation also facilitates improved estimation of molecular hydrogen density.  
In the remainder of \S\ref{sec-IV:5}, we mainly focus on the MC-RTF run in which radiation fields are tracked through transport calculation, unless specified otherwise.

\begin{figure*}[t]
\epsscale{1.15}
\plotone{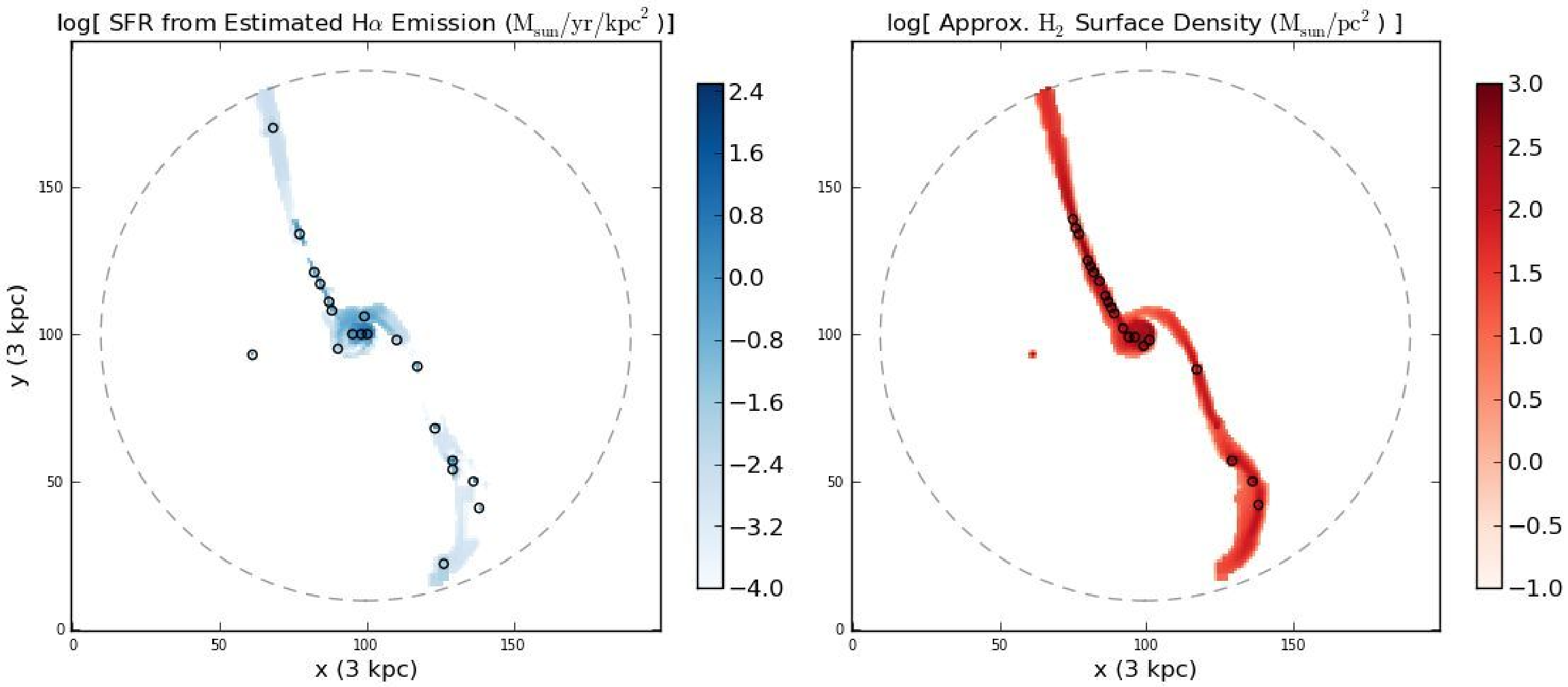}
    \caption{{\it Left:} SFR surface density, $\Sigma_{\rm SFR, \,est}$, in the unit of $M_{\odot}\,{\rm yr}^{-1}\,{\rm kpc}^{-2}$ evaluated from a mock observation of ${\rm H}\alpha$ emission using Eq.(\ref{eq:Ha_conversion}) in the MC-RTF run.  Shown at 13.3 Myr into the high-resolution evolution in a 3 kpc box centered on a clump that actively forms SFMC particles.  These face-on projections are generated with uniform 15 pc resolution.  The axes of the image indicate the number of resolution elements. Locations of 20 local maxima are marked with {\it black circles}. 
     {\it Right:} ${\rm H}_2$ surface density, $\Sigma_{\rm H_2, \,est}$, in the unit of $M_{\odot}\,{\rm pc}^{-2}$ evaluated with Eq.(\ref{eq:f_H2}).  Again, locations of 20 local maxima are marked with {\it black circles}.  Note that ${\rm H}\alpha$ and ${\rm H}_2$ peaks are closely related, but do not perfectly coincide.   
\label{fig:maps_preview}}
\end{figure*}

\subsubsection{Simulated Map of SFR Surface Density from $H\alpha$ Emission} 

To begin with, we describe how the {\it SFR surface density} is mapped from a mock observation of ${\rm H}\alpha$ emission.   
In the hot HII regions around young massive stars, the main process contributing to the Balmer lines is the recombination of ionized hydrogen, ${\rm H}^+$. 
Therefore, first and foremost, we estimate the ${\rm H}\alpha$ emission by recombination per volume element $dV$ as
\begin{eqnarray}
dL_{{\rm H}\alpha, \,{\rm R}} &=& 4 \pi j_{{\rm H}\alpha, \,{\rm R}} dV \\
&=& 4 \pi \times 2.82 \times 10^{-26} \,\,T_4^{\,-0.942-0.031\,{\rm ln}T_4} \,n_{\rm e} n_{{\rm H}^+} dV
\label{eq:Ha_by_recomb}
\end{eqnarray}
in ${\rm erg\,\, s^{-1}}$, where $j_{{\rm H}\alpha, \,{\rm R}}$ refers to the ${\rm H}\alpha$ emission rate by recombination \citep[Eq.(9) of][]{2011ApJ...727...35D}, $n_{\rm e}$ and $n_{\rm H^+}$ to the number density of e and ${\rm H}^+$, respectively, and $T_4$ to the temperature $T_{\rm e}$ in the unit of $10^4\,\, {\rm K}$.
Second, we add the enhancement of ${\rm H}\alpha$ emission by collisional excitation.  
This boost is usually minor because, in typical HII regions, the neutral hydrogen fraction is very small, and few free electrons have sufficient energy to excite bound electrons of neutral hydrogen to $n \geq 3$ \citep{2003ApJ...592..846L}. 
We estimate the contribution of collisional excitation $n=1 \rightarrow 3$ followed by de-excitation $n=3 \rightarrow 2$ as
\begin{eqnarray}
dL_{{\rm H}\alpha, \,{\rm C}} &=& E_{32} q_{13} \,n_{\rm e} n_{\rm H} dV \\
&=& E_{32} \left({2\pi \hbar^4 \over k_{\rm B}m_{\rm e}^3} \right)^{1/2} {\Gamma_{13}(T_{\rm e}) \over \omega_1 \sqrt{T_{\rm e}}} e^{-E_{31}/ k_{\rm B} T_{\rm e}} \,n_{\rm e} n_{\rm H} dV \\
&=& 1.30 \times 10^{-17} \,\,{\Gamma_{13}(T_{\rm e}) \over \sqrt{T_{\rm e}}} e^{-12.1\,{\rm eV}/ k_{\rm B} T_{\rm e}} \,n_{\rm e} n_{\rm H} dV
\label{eq:Ha_by_colexc}
\end{eqnarray}
in ${\rm erg\,\, s^{-1}}$, where $E_{ij}$ is the energy associated with the transition $i \rightarrow j$, $q_{13}$ is the rate coefficient of excitation $n=1 \rightarrow 3$, $\Gamma_{13}(T_{\rm e})$ is the Maxwellian-averaged effective collision strength,\footnote{$\Gamma_{13}(T_{\rm e}) = 0.350 - 2.62\times10^{-7}\,T_{\rm e} - 8.15\times10^{-11}\,T_{\rm e}^2+6.19\times10^{-15}\,T_{\rm e}^3$ for $4000\,\,{\rm K}<T_{\rm e}< 25000\,\,{\rm K}$ from Table 3 of \cite{1983MNRAS.202P..15A}.} $\omega_1=2$ is the statistical weight of the originating level, $k_{\rm B}$ is the Boltzmann constant, $m_{\rm e}$ is the mass of an electron, and $(2\pi \hbar^4 / k_{\rm B}m_{\rm e}^3)^{1/2} = 8.63 \times 10^{-6}$.
Here we fix $T_{\rm e}$ at 8000 K, a characteristic temperature inside a typical HII region that is poorly resolved in our simulation.\footnotemark\,
Third, we subtract the emission from the diffuse ionized gas (DIG) by removing the contribution by cells whose proton number densities are smaller than $10\,\,{\rm cm}^{-3}$ or whose temperatures higher than $10^{5.5}\,{\rm K}$ \citep{2010ApJ...722.1699S}.
Lastly, we convert the luminosities in ${\rm H}\alpha$ emission into the star formation rates with
\begin{eqnarray}
dM_* / dt \,\,(M_{\odot}\,\,{\rm yr}^{-1}) = 5.3 \times 10^{-42}\,\, L_{{\rm H}\alpha}\,\,({\rm erg\,\, s^{-1}})
\label{eq:Ha_conversion}
\end{eqnarray}
in which a correction for dust extinction is not considered \citep[e.g.][]{2007ApJ...666..870C}.  
\footnotetext{Collisional excitation may falsely dominate ${\rm H}\alpha$ luminosities if the temperature is inadequately evaluated because Eq.(\ref{eq:Ha_by_colexc}) has a steep temperature dependence. 
In fact, individual HII regions in the reported simulation are not properly resolved, and monochromatic photons (16.0 eV) from SFMC particles often keep overly-heated temperature ($> 2\times10^4\,\,{\rm K}$) in the neighboring under-resolved HII regions.    
When this temperature is naively adopted, Eq.(\ref{eq:Ha_by_colexc}) might overestimate $L_{{\rm H}\alpha, \,{\rm C}}$. 
In an attempt to compensate for our poor resolution, the temperature in Eq.(\ref{eq:Ha_by_colexc}) is thus fixed at 8000 K, a characteristic temperature inside a typical HII region.    
With this correction, the contribution by collisional excitation is indeed found to be only marginally important.}

In order to distinguish the SFR densities measured in these mock observations from the actual values in simulations (e.g. $\Sigma_{\rm SFR}$ in \S\ref{sec-IV:4-SF}), we hereafter define $\rho_{\rm SFR,\,est} = 5.3 \times 10^{-42}\,dL_{{\rm H}\alpha}/dV$ and $\Sigma_{\rm SFR,\,est}$ as the SFR (surface) density estimated from a simulated observation.  
In Appendix \ref{sec:appendix-A}, by comparing the ``estimated'' SFR surface density from ${\rm H}\alpha$ emission, $\Sigma_{\rm SFR,\,est}$, with the ``actual'' SFR surface density, $\Sigma_{\rm SFR}$, using a 1.2 kpc resolution mock observation, we find that our ${\rm H}\alpha$ emission estimate, Eq.(\ref{eq:Ha_by_recomb})$+$(\ref{eq:Ha_by_colexc}), and our adopted ${\rm H}\alpha$-to-SFR conversion factor, Eq.(\ref{eq:Ha_conversion}), are reliable.

\begin{figure}[t]
\epsscale{1.16}
\plotone{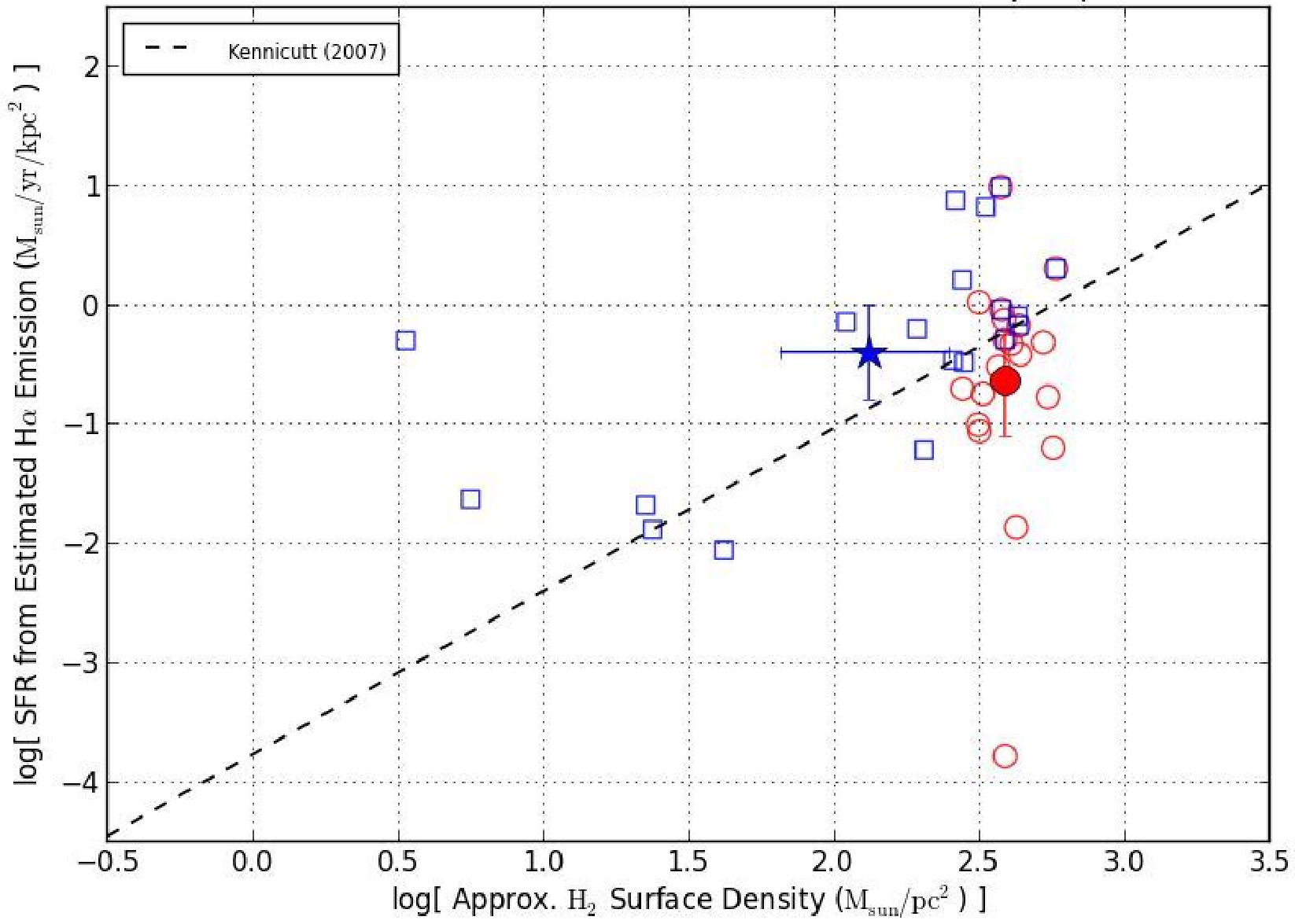}
    \caption{Distribution of local maxima found in Figure \ref{fig:maps_preview} on the plane of SFR surface density, $\Sigma_{\rm SFR,\,est}$ (from mock ${\rm H}\alpha$ emission), and approximate ${\rm H}_2$ surface density, $\Sigma_{\rm H_2,\,est}$.   Peaks of $\Sigma_{\rm SFR,\,est}$ are shown as {\it blue squares} and peaks of $\Sigma_{\rm H_2,\,est}$ {\it red circles}.  A {\it blue filled star} represents the average $\Sigma_{\rm SFR,\,est}$ ($y$-axis) and the average $\Sigma_{\rm H_2,\,est}$ ($x$-axis) for the peaks in the $\Sigma_{\rm SFR,\,est}$ map, while a {\it red filled circle} is for the $\Sigma_{\rm H_2,\,est}$ peaks.  These markers are chosen to mimic Figure 4 of \cite{2010ApJ...722.1699S}.  Error bars represent 95\% confidence intervals from $10^4$ bootstrap samples.  Plotted as a dashed line is the best fit to the spatially-resolved observation of M51 by \cite{2007ApJ...671..333K}.  
\label{fig:KS_preview}}
\end{figure}

\begin{figure*}[t]
\epsscale{1.17}
\plotone{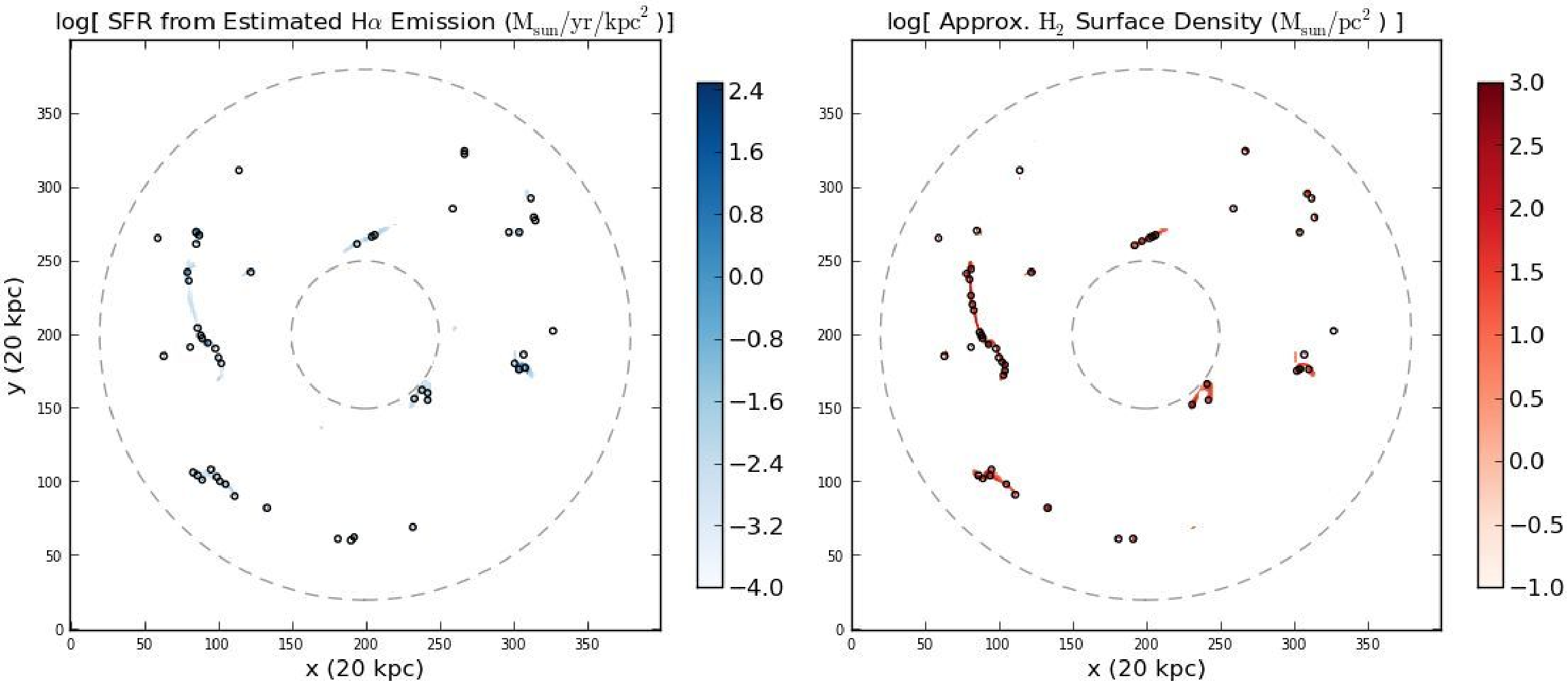}
    \caption{Estimated SFR surface density ({\it left}, in $M_{\odot}\,{\rm yr}^{-1}\,{\rm kpc}^{-2}$, from mock ${\rm H}\alpha$ emission) and ${\rm H}_2$ surface density ({\it right}, in $M_{\odot}\,{\rm pc}^{-2}$).   Shown at 13.3 Myr into the high-resolution evolution of the MC-RTF run in a 20 kpc box centered on the galactic center.  The same mock observation technique as in Figure \ref{fig:maps_preview} is employed with uniform 50 pc resolution.  The central region of radius 2.5 kpc is cut out in these maps (inner dashed circle; see \S\ref{sec-IV:2} and \S\ref{sec-IV:5-spatial}).  The locations of 50 local maxima are marked with {\it black circles}.  Note that star-forming clumps are highly clustered; compare with the projected density in the left panel of Figure \ref{fig:gammaHI_eVscm3_up}.  
\label{fig:maps_50pc}}
\end{figure*}

\subsubsection{Simulated Map of $H_2$ Surface Density} 

We now discuss how the {\it $H_2$ surface density} is mapped.  
Molecular hydrogen is not explicitly traced in our chemistry solver which follows only six atomic species.  
We instead use an approximation to estimate the molecular hydrogen fraction, $f_{\rm H_2}$, as a function of gas density and metallicity by
\begin{eqnarray}
f_{\rm H_2} \simeq 1-{3 \over 4}{s \over1+0.25\,s} 
\label{eq:f_H2}
\end{eqnarray}
with parameters defined as
\begin{eqnarray}
s &=& {{\rm ln}(1+0.6\,\chi+0.01\,\chi^2) \over 0.6\, \tau_{\rm c}}, \\
\chi &\simeq& 0.77 \,(1+3.1\,\, Z\,'^{\,\,0.365} ), \\
\tau_{\rm c} &=& {\Sigma \sigma_{\rm d} \over \mu_{\rm H} } \,\,\,\simeq\,\,\, 321 \,\,Z\,' {\rho_{\rm gas}^2 \over |\nabla \rho_{\rm gas}|}, 
\end{eqnarray}
where approximations in \cite{2012ApJ...749...36K} have been applied to the original formula of the scaled radiation field $\chi$ \citep{2008ApJ...689..865K, 2009ApJ...693..216K, 2010ApJ...709..308M}, and an approximated column density of the cell, $\Sigma = \rho^2/|\nabla \rho|$, has been employed for the cell's dust optical depth $\tau_{\rm c}$  \citep[the Sobolev approximation described in][]{2011ApJ...729...36K}.
Here $Z\,'$ is the metallicity normalized to the solar value, $\sigma_{\rm d}$ is the mean extinction cross section by dust per hydrogen nucleus, $\mu_{\rm H}$ is the mean mass per hydrogen nucleus, and $\rho_{\rm gas}$ is in the unit of ${\rm g\,cm^{-3}}$. 

We hereafter define $\rho_{\rm H_2,\,est}=f_{\rm H_2}\rho_{\rm gas}$ and $\Sigma_{\rm H_2,\,est}$ as the ${\rm H}_2$ (surface) density evaluated from a simulated galaxy.  
This is to emphasize that the our ${\rm H}_2$ density is an approximate estimate, not a real value which is explicitly followed in simulations.  
In Appendix \ref{sec:appendix-B} we discuss whether the use of the \cite{2008ApJ...689..865K, 2009ApJ...693..216K} equilibrium model is adequate in our experiment to determine molecular gas content.
We also note that in observational studies, $^{12}$CO $J = 1\rightarrow0$ transition line emission is typically used to trace molecular gas, and we are neglecting any additional scatter in the SFR relation that might be induced by spatially-variable CO-to-H$_2$ conversion factors. 
The observations of the spatially-resolved star formation law to which we will compare below all target normal spiral galaxies of roughly solar metallicity, and both observations and theoretical investigations of the CO-to-H$_2$ conversion factor in such galaxies suggest that its variation is at the factor of $\sim 2$ level \citep[e.g.][]{2010ApJ...716.1191W, 2011MNRAS.418..664N, 2012MNRAS.421.3127N, 2012ApJ...747..124F, 2012ApJ...758..127F, 2013arXiv1301.3498B}. 
Thus the error we make by neglecting this aspect of the observations is likely to be a fairly modest effect.

\subsubsection{Correlation Between SFR and $H_2$ Surface Densities} 

Because the location of ionized gas is self-consistently computed in our high-resolution radiation hydrodynamics simulation, we are able to investigate the correlation between ionized and molecular gas in detail.  
With $\rho_{\rm SFR, \,est}$ and $\rho_{\rm H_2,\,est}$ estimated as above, projections of both densities can be readily generated. 
One such example is shown in Figure \ref{fig:maps_preview} where a face-on projection of SFR density, $\Sigma_{\rm SFR, \,est}$, and that of ${\rm H}_2$ density, $\Sigma_{\rm H_2, \,est}$, are produced  and compared side by side.
These maps are generated with uniform 15 pc resolution\footnotemark\, at 13.3 Myr into the high-resolution evolution of the MC-RTF run, in a 3 kpc box centered on a gas clump that harbors multiple young SFMC particles.
\footnotetext{Using {\tt FixedResolutionBuffer} in {\it yt} \citep{2011ApJS..192....9T} that deposits meshes of varying sizes into a pixelized, fixed-resolution array.}
Also displayed are the locations of 20 local maxima of each map marked with black circles.\footnote{Using {\tt maximum\_filter} in scipy package  at http://www.scipy.org/.}  
The resolution of these maps can be regarded as a mock observation equivalent of an aperture size (diameter) in photometric observations of local galaxies. 
For Figure \ref{fig:maps_preview}, the resolution of 15 pc is chosen in order to demonstrate how the locations of peaks in one map differ from the ones in the other. 
In these images, it is unmistakable that ${\rm H}\alpha$ peaks and ${\rm H}_2$ peaks are closely related, but they do not always coincide perfectly.   
This experiment verifies the proposition we discussed in the beginning of this section.
That is, the site of highly ionized gas (typically identified with ${\rm H}\alpha$ peaks) may not necessarily overlap with the site of on-going star formation (typically identified with ${\rm H}_2$ peaks).  

\begin{figure*}[t]
\epsscale{1.14}
\plotone{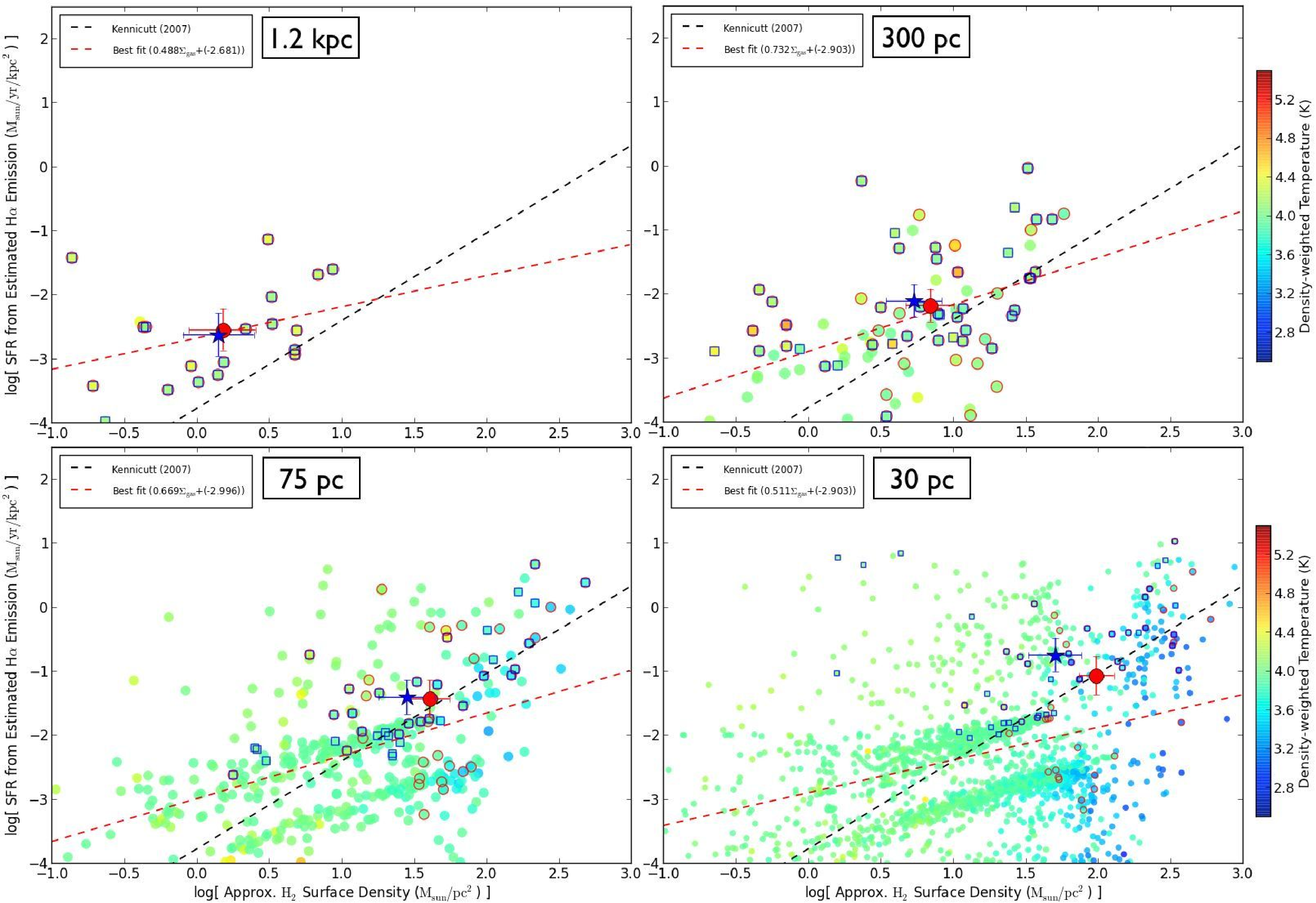}
    \caption{The distribution of pixels of galaxy maps (e.g. Figure \ref{fig:maps_50pc}) on the plane of SFR surface density (from mock ${\rm H}\alpha$ emission) and ${\rm H}_2$ surface density.  Each panel renders 1200, 300, 75, and 30 pc aperture averages at 13.3 Myr into the high-resolution evolution.  Points are colored by projected temperature weighted by density.  Also displayed are local peaks in $\Sigma_{\rm SFR,\,est}$ ({\it blue squares}) and $\Sigma_{\rm H_2,\,est}$ ({\it red circles}) as in Figure \ref{fig:KS_preview}.  A {\it blue filled star} and a {\it red filled circle} each represents the average location of each set of peaks.  Error bars indicate 95\% confidence intervals from $10^4$ bootstrap samples.  A {\it red dashed line} shows a best fit in each panel, while a {\it black dashed line} denotes a best observational fit by \cite{2007ApJ...671..333K}.  The color version of this figure is available in the electronic edition.  
    \label{fig:KS_multiple}}
\end{figure*}

To portray this discrepancy in a clearer way, in Figure \ref{fig:KS_preview} we place the local maxima of Figure \ref{fig:maps_preview} on the plane of SFR surface density, $\Sigma_{\rm SFR,\,est}$, and ${\rm H}_2$ surface density, $\Sigma_{\rm H_2,\,est}$. 
Both axes represent estimated values from the simulated galaxy, MC-RTF, at 13.3 Myr into the high-resolution evolution. 
Each data point can be regarded as a 15 pc aperture average at the local maxima in the map of either $\Sigma_{\rm SFR,\,est}$ or $\Sigma_{\rm H_2,\,est}$.  
SFR surface density peaks are marked with blue circles and ${\rm H}_2$ peaks with red circles.
It is obvious that the $\Sigma_{\rm SFR,\,est}$ peaks are not necessarily the peaks in $\Sigma_{\rm H_2,\,est}$ map, and vice versa.  
Further, some data points suggest that $\Sigma_{\rm SFR,\,est}$ peaks could have very low $\Sigma_{\rm H_2,\,est}$ (upper left in the panel), and vice versa (lower right in the panel).  
In other words, the locations of highly ionized gas do not precisely correlate with those of dense molecular gas.
The discrepancy can be encapsulated by a separation between the averages of the two sets of peaks.  
A blue star represents the average $\Sigma_{\rm SFR,\,est}$ and $\Sigma_{\rm H_2,\,est}$ for the peaks of the $\Sigma_{\rm SFR,\,est}$ map, whereas a red filled circle represents the average for the peaks of the $\Sigma_{\rm H_2,\,est}$ map.  
The separation between the blue star and the red filled circle characterizes how well ${\rm H}\alpha$ performs as star formation tracers.  
The separation becomes particularly large at high resolution, i.e. small aperture size. 
We will further discuss this issue later in \S\ref{sec-IV:5-spatial}.  

\subsection{Spatially-resolved Star Formation Relation}  \label{sec-IV:5-spatial}

We now apply the mock observation techniques described in \S\ref{sec-IV:5-obs} to the entire galactic disk.   
With these techniques, it is possible to directly compare star formation tracers between simulated galaxies and observed ones.  
We can also observe how stellar feedback manifests itself in the correlation between ionized and molecular gas at various resolutions. 

Figure \ref{fig:maps_50pc} shows the same estimation of SFR surface density and ${\rm H}_2$ surface density, as in Figure \ref{fig:maps_preview}.  
Instead of focusing on a single star-forming clump, Figure \ref{fig:maps_50pc} looks at the galactic disk of radius 9 kpc (outer dashed circle) that encompasses most of the active SFMC particles.
The uniform resolution of these maps, 50 pc, can be viewed as a diameter of apertures in our simulated observation. 
As in Figure \ref{fig:gammaHI_eVscm3_up}, the central region of radius 2.5 kpc from the galactic center is specified with a small dashed circle.  
We exclude the contribution by this region because the SFMC particles there have not emitted ionizing photons by design (see \S\ref{sec-IV:2}) and may thus contaminate our analysis.   
The locations of 50 local maxima are marked with black circles, which tend to cluster in a few small areas.
To be defined as local maxima, peaks should be away from one another by at least 4.5 resolution elements.
This mock observation carefully mimics the highest resolution observation of a local galaxy, M33, in \citet[][compare with their Figure 1]{2010ApJ...722.1699S}.
From our high-resolution galaxy data, we can also make galactic maps like Figure \ref{fig:maps_50pc} with arbitrary resolution (aperture size) as long as it is larger than the simulation's spatial resolution, 3.8 pc.  
In the subsequent analysis, we analyze six additional maps generated at different resolutions: 1200, 600, 300, 150, 75, and 30 pc.
To match the observational procedure by \cite{2010ApJ...722.1699S}, for a series of these resolutions we find up to 50 local maxima that are nearest to the peak locations at 50 pc resolution.\footnote{They first identified CO regions and ${\rm H}\alpha$ peaks using their smallest aperture, $d=75$ pc.  Then an aperture of diameter [1200, 600, 300, 150, 75 pc] is centered on each CO and ${\rm H}\alpha$ peak to measure the fluxes within that aperture.  The authors however have found that the qualitative results in the subsequent sections (such as Figure 9) are independent of the peak finding algorithm.}

We then distribute all the pixelized elements of each galaxy map on the plane of SFR surface density and ${\rm H}_2$ surface density.
The plots resulting from this process are in Figure \ref{fig:KS_multiple}, where we show how the pixels on a galaxy map are spread with respect to their SFR surface densities and ${\rm H}_2$ surface densities.
The panels represent four different resolutions of the map, and each of the data points is colored by density-weighted projected temperature.  
Data points with highlighted edges stand for local peaks in $\Sigma_{\rm SFR,\,est}$ (blue squares) and in $\Sigma_{\rm H_2,\,est}$ (red circles).
A blue star and a red filled circle each represents the average location of each set of peaks. 
The error bars here indicate 95\% confidence intervals from $10^4$ bootstrap samples. 

\begin{figure*}[t]
\epsscale{1.15}
\plotone{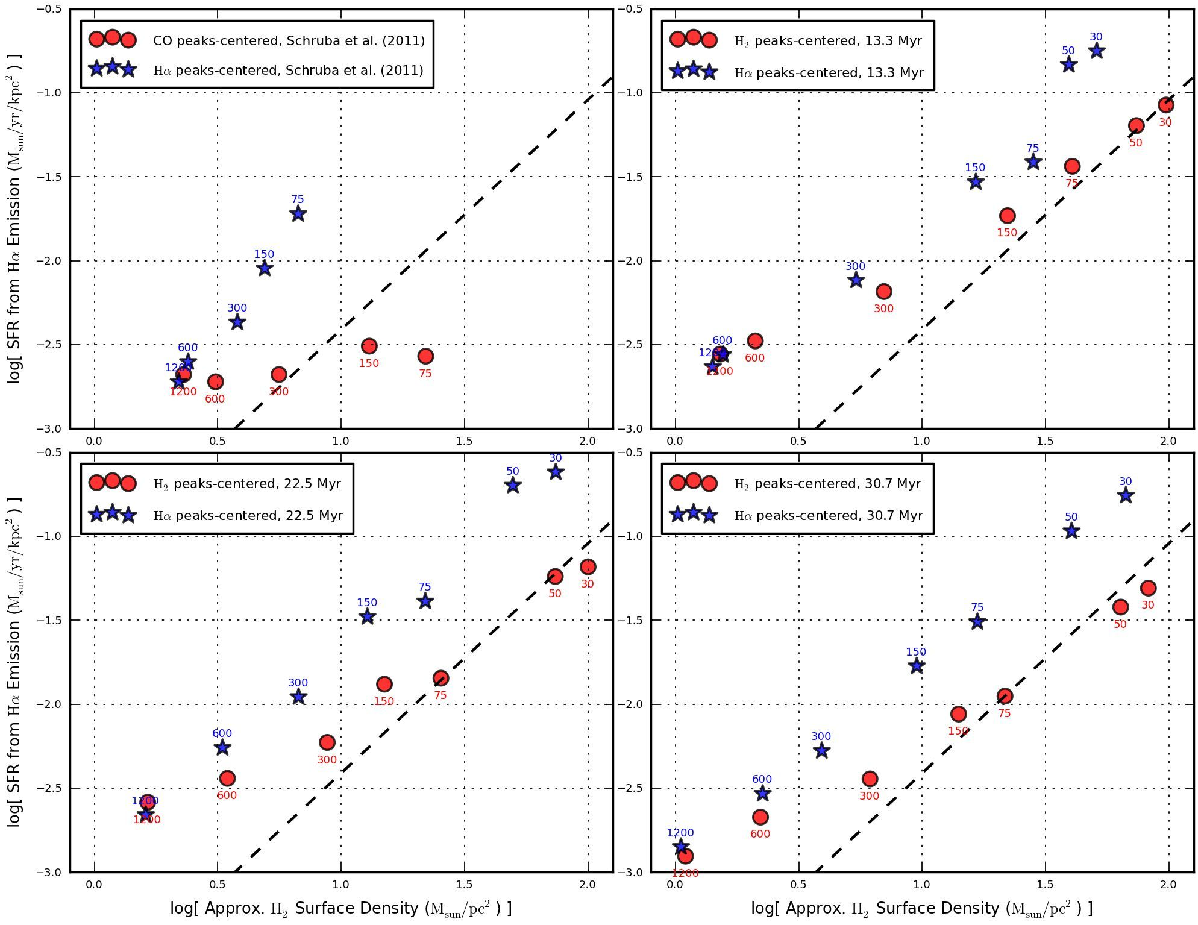}
    \caption{{\it Top left:}  {\it blue stars} show the median  $\Sigma_{\rm SFR}$ and $\Sigma_{\rm H_2}$ for apertures centered on ${\rm H}\alpha$ peaks of M33.  {\it Red filled circles} show the same for apertures centered on CO peaks.  Each point is annotated by its aperture size.  A {\it dashed line} denotes a best fit to M51 by \cite{2007ApJ...671..333K}.  Adapted from Figure 4 of \cite{2010ApJ...722.1699S}.  {\it Top right:} the simulation equivalent of the top panel in our MC-RTF run at 13.3 Myr into high-resolution evolution.  Average locations of local maxima found in Figure \ref{fig:KS_multiple} are displayed.  Each point is annotated by the resolution of the performed mock observation.   {\it Bottom left:} same as the top right panel but at 22.5 Myr into high-resolution evolution.  {\it Bottom right:} same as the top right panel but at 29.9 Myr into high-resolution evolution.
\label{fig:Schruba_fig4}}
\end{figure*}

When compared with observations, Figure \ref{fig:KS_multiple} provides us with an illuminating view of the galactic star formation. 
First and foremost, one of the most unambiguous features in these panels is the large scatter in the data, up to $\pm3$ orders of magnitude away from the best fit found in M51 using a 520 pc aperture \citep[black dashed line;][]{2007ApJ...671..333K}, especially in the mock observations with high resolutions, $\lesssim$ 75 pc. 
The star formation relation observed on $\sim$ kpc scales seem to rapidly break down once aperture sizes reach $\lesssim$ 75 pc that is comparable to the scales of GMCs.
This result is consistent with the recent observations of local galaxies by many authors using smaller aperture sizes down to 75 pc \citep[e.g.][]{2010ApJ...722L.127O, 2010ApJ...722.1699S,  2011ApJ...735...63L}.

Each data point in these plots averages over star-forming regions with a range of evolutionary stages  and diverse ${\rm H}\alpha$-to-${\rm H}_2$ ratios \citep[e.g.][]{2010ApJ...722.1699S}.  
Therefore, a data point with a large aperture \citep[$\gtrsim$ 300 pc; e.g.][]{2007ApJ...671..333K} or a disk-averaged observation \citep[e.g.][]{1998ApJ...498..541K} yields not only a {\it spatially-averaged} estimate, but also a {\it temporally-averaged} one.
In contrast, a data point with a small aperture ($\lesssim$ 75 pc) captures only particular stages of GMC evolution, because the aperture is too small to average over a wide range of the evolutionary stages of GMCs.  
Hence, the correlation between ionized and molecular gas becomes loose at small scales. 
To further illustrate this effect, we paint various evolutionary stages of each data point (star-forming region) in Figure \ref{fig:KS_multiple} by its projected temperature. 
In particular, in the 30 pc resolution panel, a mild but clear transition is pronounced from lower right to upper left.  
It indicates a wide spectrum of star-forming regions on a single galaxy, from a cold dense region that actively forms stars (lower right) to a hot diffuse region that is likely dominated by the radiation from neighboring star clusters (upper left; hence exhibiting less active star formation).  
This supports the idea that at a size comparable to the scales of GMCs, each aperture represents only a small fraction of evolutionary stages of GMCs \citep{2010ApJ...722L.127O, 2010ApJ...722.1699S,2012ApJ...757..138K}.
We will re-investigate this issue in \S\ref{sec-IV:5-environ}.\footnote{Readers may notice that the data points in the bottom panel of Figure \ref{fig:KS_multiple} are grouped into two prominent sequences. This occurs because there are a large number of SFMC particles associated with two prominent clumps in the simulation, both relatively near the galactic center, that are systematically offset from each other in the Kennicutt-Schmidt plane. The reason for this offset will be investigated in future work.} 

\begin{figure*}[t]
\epsscale{1.13}
\plotone{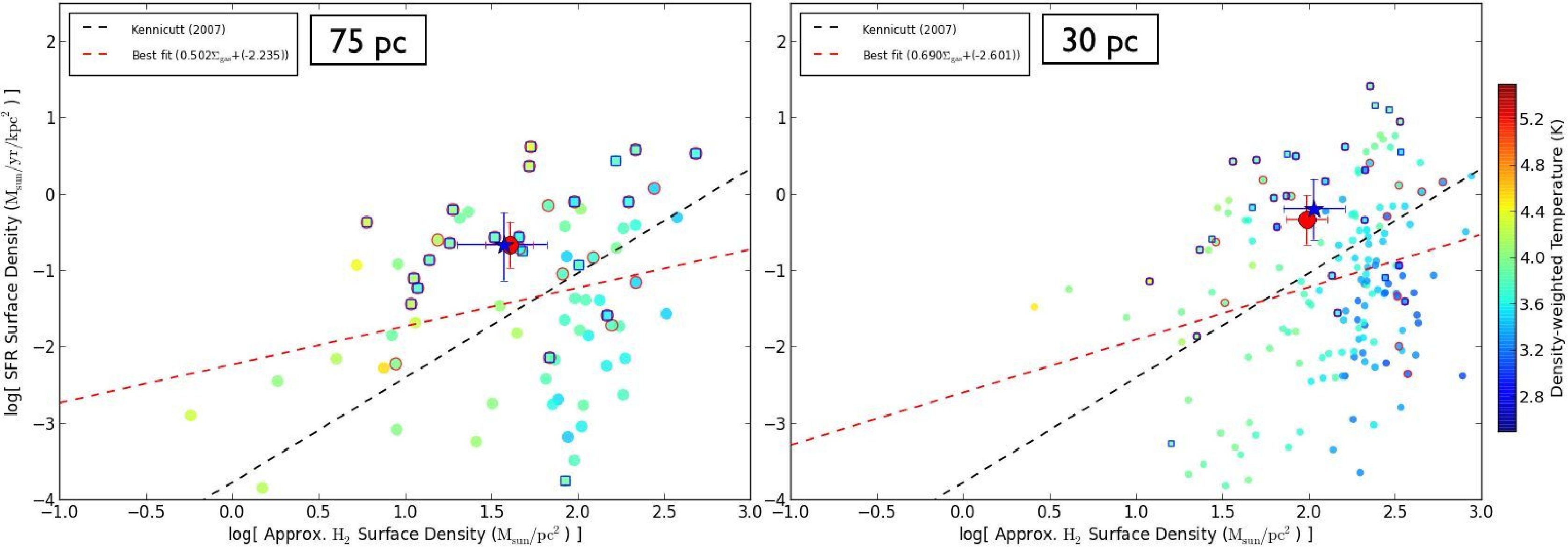}
    \caption{An analogue of bottom panels of Figure \ref{fig:KS_multiple} using the ``actual'' SFR surface density in $y$-axis (not the ``estimated'' SFR from the mock ${\rm H}\alpha$ emission) and the ${\rm H}_2$ surface density in $x$-axis.  Each panel renders 75 and 30 pc aperture averages at 13.3 Myr into the high-resolution evolution.  See the caption of Figure \ref{fig:KS_multiple} for a detailed description of these plots.  The color version of this figure is available in the electronic edition.
\label{fig:Halpha_SFR_compare_2}}
\end{figure*}

Finally, we focus on the locations and averages of $\Sigma_{\rm SFR,\,est}$ and $\Sigma_{\rm H_2,\,est}$ peaks, using the same technique and notation described in Figure \ref{fig:KS_preview}.  
The top right panel of Figure \ref{fig:Schruba_fig4} shows the average locations of up to 50 local maxima identified at seven different resolutions.  
It is the simulation equivalent of the similar analysis performed by \citet[][top left panel of Figure \ref{fig:Schruba_fig4}]{2010ApJ...722.1699S}.
As was examined in \S\ref{sec-IV:5-obs} and Figure \ref{fig:KS_preview}, the separation between the blue star and red circle encapsulates the discrepancy between the highly ionized gas and dense molecular gas at a given resolution.  
It is evident that the separation between the two averages grows in both panels with smaller aperture size, i.e. higher resolution.
This is because at high spatial resolution, an aperture at a ${\rm H}\alpha$ peak tends not to enclose the star-forming regions in cold dense clumps, whereas an aperture at a ${\rm H}_2$ peak tends not to include the star-forming regions in hot diffuse environment (more to be discussed in \S\ref{sec-IV:5-environ}).   
Note also that the offset is consistently observed from 13.3 Myr to 30.7 Myr into high-resolution evolution (bottom panels), which indicates that the observed trend is not caused by a galaxy at a peculiar moment.

Consequently, one may conclude that ${\rm H}\alpha$ emission is not the best star formation indicator at resolutions of $\lesssim$ 75 pc.   
It implies that the spatially-resolved star formation relation in local galaxies could potentially be influenced by observation strategies (e.g. which peaks to choose, which aperture sizes to use).  
We note that once we adopt the ``actual'' value of SFR recorded in the simulation instead of its tracer ${\rm H}\alpha$, the mismatch between SFR and ${\rm H}_2$ peaks at high resolution disappears.
In Figure \ref{fig:Halpha_SFR_compare_2} we plot an analogue of Figure \ref{fig:KS_multiple} by using the ``actual'' SFR surface density, $\Sigma_{\rm SFR}$, recorded in simulations (see Appendix \ref{sec:appendix-A}) instead of the ``estimated'' one, $\Sigma_{\rm SFR,\,est}$, from the mock ${\rm H}\alpha$ emission.  
When compared with the bottom panels of Figure \ref{fig:KS_multiple}, one can immediately notice that the mismatch between $\Sigma_{\rm SFR}$ and $\Sigma_{\rm H_2}$ peaks becomes negligible even at these high resolutions. 
The average locations of peaks (a blue star and a red filled circle) also overlap.
This exercise reaffirms the idea that the displacement between SFR and ${\rm H_2}$ peaks observed in \S \ref{sec-IV:5-spatial} is largely because we measure SFR with ${\rm H}\alpha$ emission.

\begin{figure*}[t]
\epsscale{0.88}
\plotone{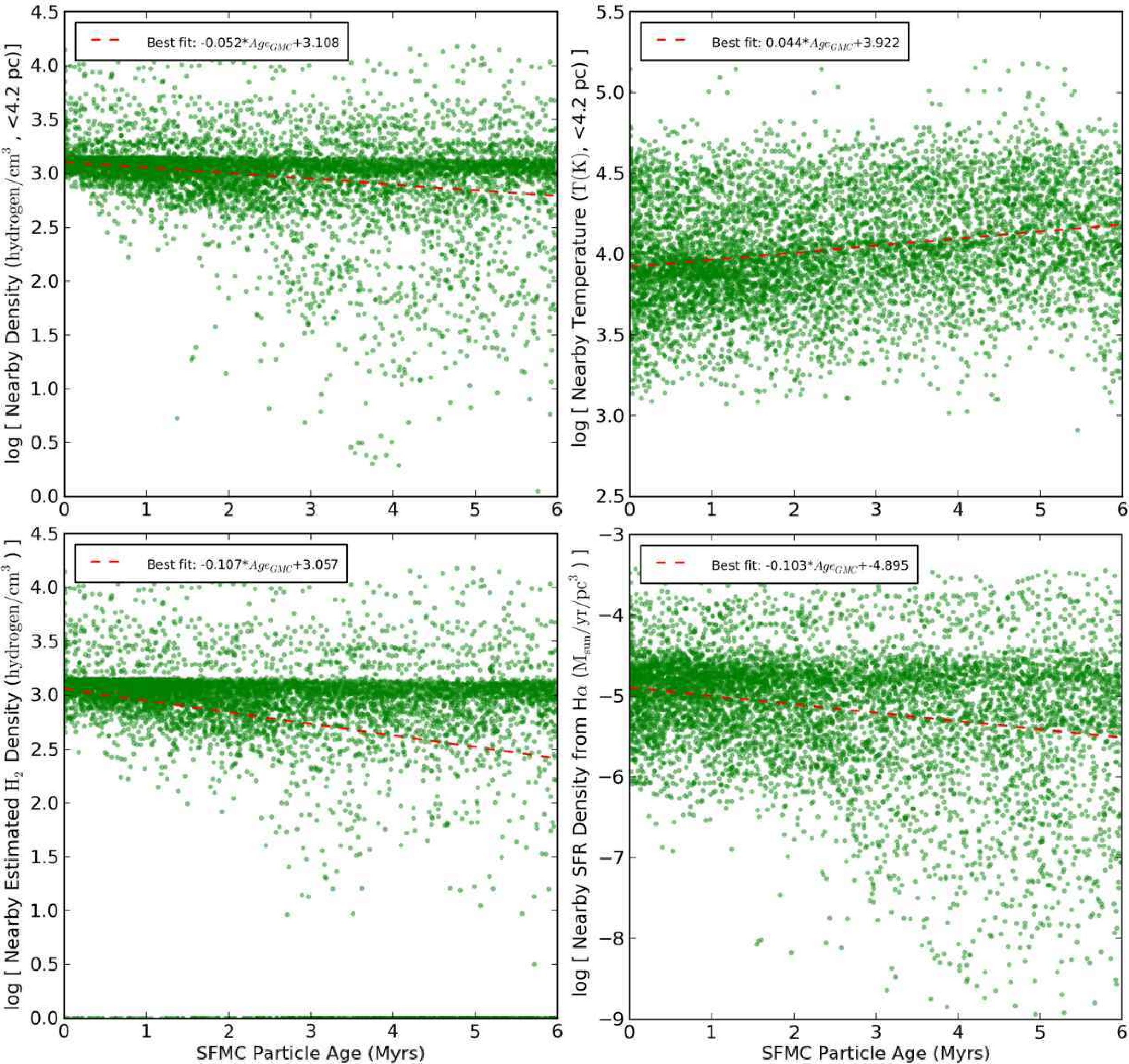}
    \caption{Gas density, temperature, ${\rm H}_2$ density, and SFR density from mock ${\rm H}\alpha$ emission in the spheres of 8.4 pc diameter centered on active SFMC particles, as functions of particle age.  For ${\rm H}_2$ density a lower bound of 1 ${\rm cm}^{-3}$ is imposed.  A {\it red dashed line}  represents a best fit in each log-normal plane.  
    \label{fig:evolution_4pc_rad}}
\end{figure*}

We also caution the readers that the offset between ${\rm H}\alpha$ peaks-centered and ${\rm H}_2$ peaks-centered averages is smaller than what is found in \cite{2010ApJ...722.1699S}.  
This may indicate that the numerical estimates of ${\rm H}\alpha$ and $f_{{\rm H}_2}$ may not be complete (see \S\ref{sec-IV:5-obs} and Appendix \ref{sec:appendix-A}).
It could also be attributed to differences between M33 and our simulated galaxy, including the difference in evolutionary stage and mass. 
Lastly, it may mean that the simulation is missing physical processes that are responsible for such a wide offset, or the numerical accuracy needed to resolve such processes. 
Indeed, we conjecture that the reason that the offset is too small is the same as the reason that our galaxy's global SFR is too high (see \S\ref{sec-IV:4-SF}); that is, our stellar feedback is inadequately week. 
Stronger feedback, particularly supernovae, would disperse clouds more quickly, and thus would likely lead to both a reduction in the overall star formation rate and a greater spread between H$\alpha$ and H$_2$ peaks. 
The converse of this point is that the spatially-resolved star formation law is a very sensitive diagnostic of how well a particular feedback recipe performs, and should be used as a test of future spatially-resolved simulations.

\begin{figure*}[t]
\epsscale{0.91}
\plotone{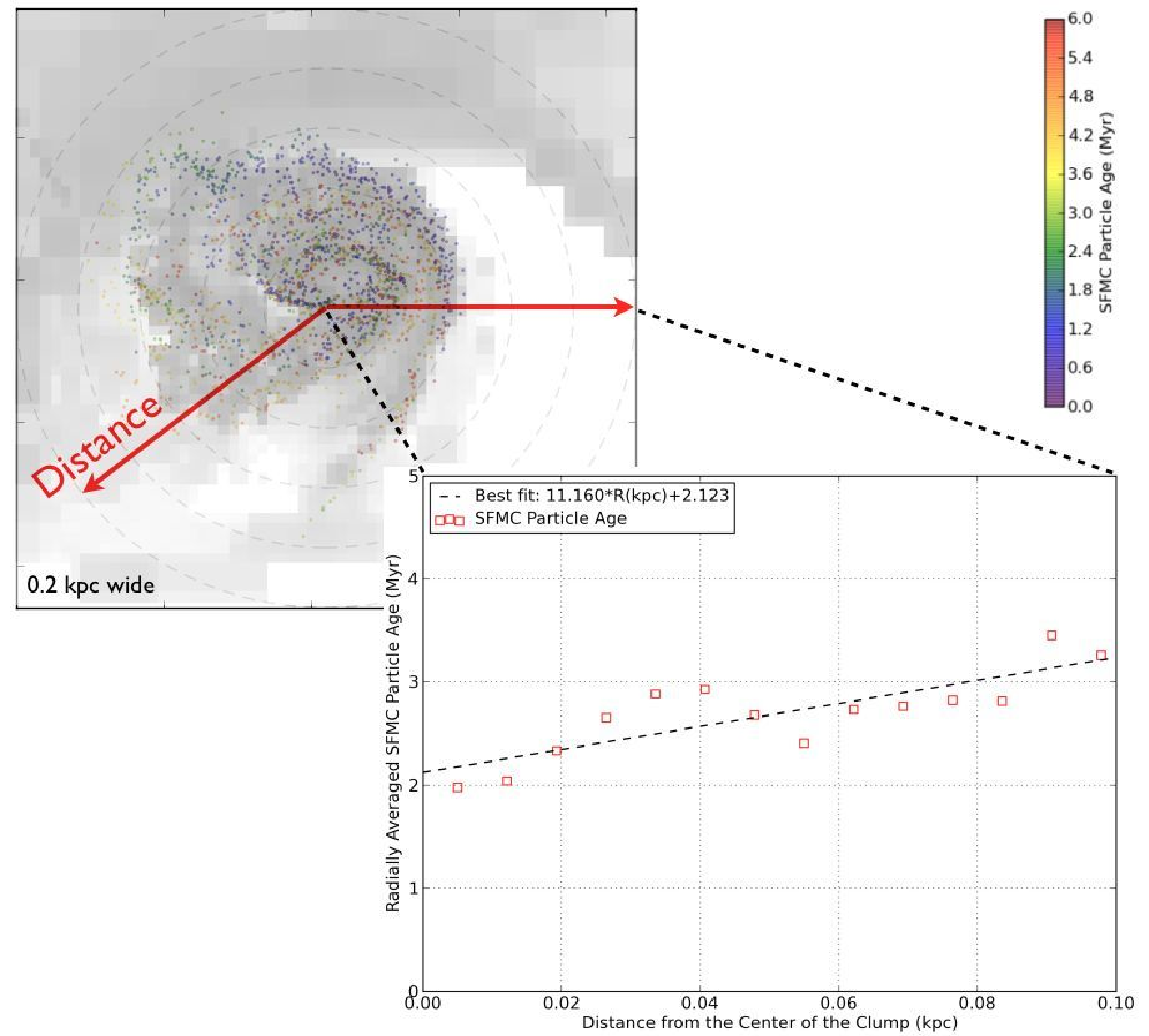}
    \caption{{\it Top feft:} a zoomed-in star-forming region  in a 0.2 kpc box.  It displays the locations of SFMC particles colored by their ages at 13.3 Myr into the high-resolution evolution, plotted on top of a black-and-white face-on projection of SFR surface density from mock ${\rm H}\alpha$ emission.  
    {\it Bottom right:} radially averages of the ages of SFMC particles as a function of distance from the center of the star-forming clump.  A {\it dashed line} denotes the best fit to these data points.  
    \label{fig:clump_radius_age}}
\end{figure*}

To summarize, we have shown in this section that the spatially-resolved Kennicutt-Schmidt relation observed at $\sim$ kpc scales breaks down at small scales comparable to the size of GMCs, $\lesssim$ 75 pc. 
The correlation varies systematically depending on whether apertures are centered on ${\rm H}\alpha$ or ${\rm H}_2$ peaks.
It is because an aperture of a GMC size captures only particular stages of GMC evolution, and because ${\rm H}\alpha$ traces hot gas around star-forming regions and is displaced from the ${\rm H}_2$ peaks themselves.  
A star formation relation of the Kennicutt-Schmidt type is very useful in depicting the globally-averaged star formation, but may not be as effective in describing every single spatially-resolved star formation events in different clumps and galaxies.  
A corollary to this statement is that one cannot correctly simulate high-resolution galaxies if the star formation is simply modeled with the Kennicutt-Schmidt relation observed at $\sim$ kpc scales. 

\subsection{Evolving Environment Around SFMC Particles}  \label{sec-IV:5-environ}

In order to understand various stages and forms of actively star-forming regions, we now turn our attention to the environment around SFMC particles.  
In Figure \ref{fig:evolution_4pc_rad}, four different physical properties in the spheres of 8.4 pc diameter\footnote{Or the smallest sphere that encloses the cell a particle belongs to.  It is almost always a sphere of 8.4 pc except the occasional cases in which particles have migrated to a coarsely refined grid.} centered on all active SFMC particles are drawn as functions of particle age, $0 < T < 6$ Myr (see \S\ref{sec-IV:2}).  
They include: proton number density, gas temperature, proton number density in ${\rm H}_2$, and SFR density from mock ${\rm H}\alpha$ emission.  
The data is from a snapshot at 13.3 Myr into the high-resolution evolution.  
For ${\rm H}_2$ density, the value of which is occasionally negligible in some cells, a lower bound of 1 ${\rm cm}^{-3}$ is imposed for a numerical purpose.  
A red dashed line represents a best fit in each log-normal plane.  

In these panels, one may immediately notice the diverse environment around SFMC particles, not only throughout their lifetime, but even at a fixed epoch of $T=0$ Myr.  
In other words, even at their birth, the (${\rm H}_2$) densities around SFMC particles manifest a large scatter up to an order of magnitude.
It is because the stellar radiation from an earlier generation of SFMC particles may heat up the star-forming gas clumps, and prevent hot dense gas of $> 10^3 \,\,{\rm cm}^{-3}$ from turning into stars. 
The scatter in (${\rm H}_2$) densities grow larger with particle age, reaching three orders of magnitude at $T=6$ Myr.
One may even observe small but meaningful evolutions of these quantities as functions of particle age, namely: decreasing (${\rm H}_2$) density and increasing temperature in time around SFMC particles.
Very young SFMC particles may likely be still enshrouded in cold dense clumps in which they were born; however, SFMC particles could soon start to drift away from their birthplaces \citep[e.g.][]{2010ApJ...722L.127O}.
This could cause the density around a SFMC particle to decrease, and the temperature to rise as the particle ages. 

\begin{figure*}[t]
\epsscale{0.88}
\plotone{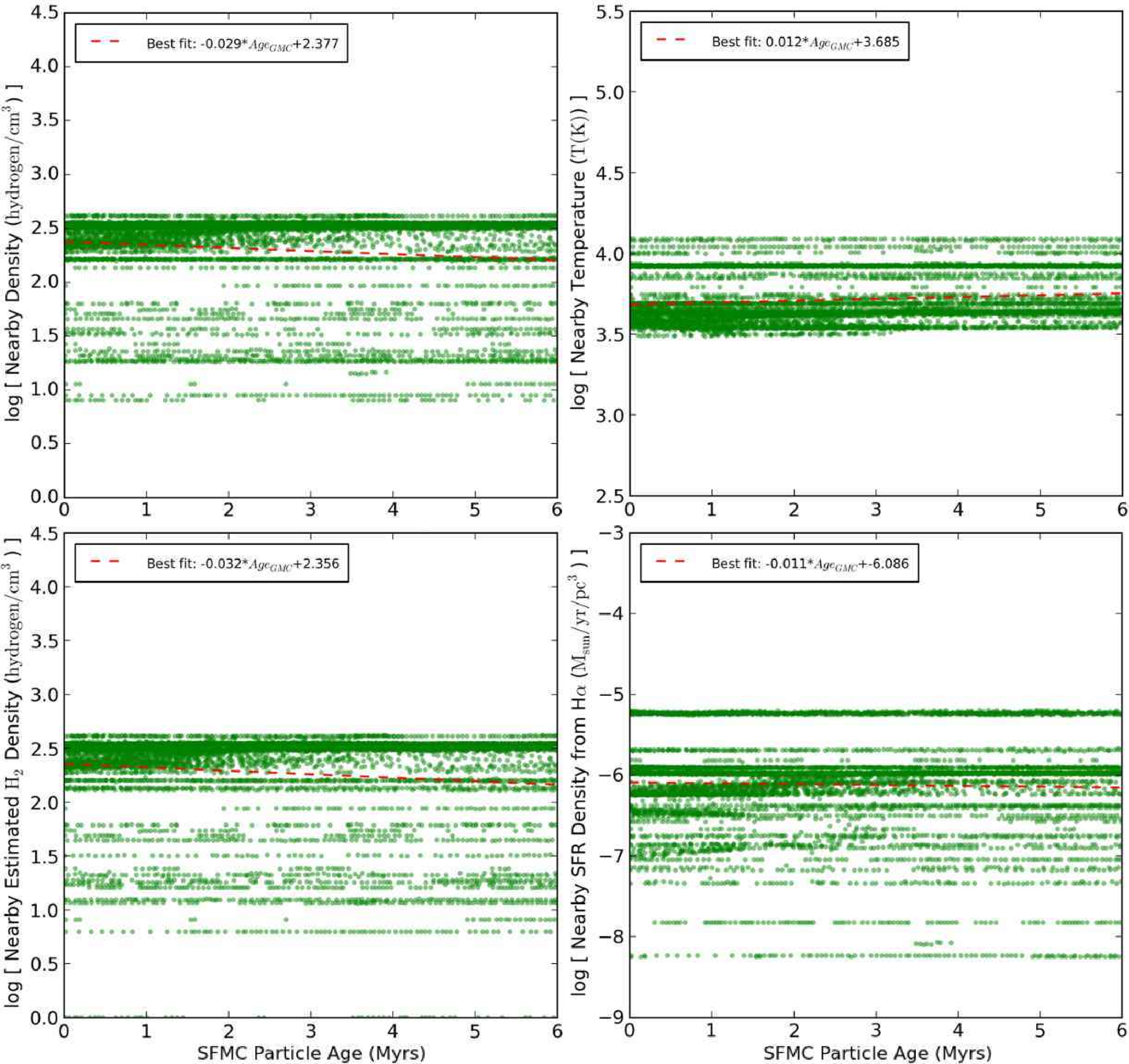}
    \caption{Same as Figure \ref{fig:evolution_4pc_rad}, but averaged in spheres of 600 pc diameter.  Because most individual star-forming regions are much smaller than the sphere of 600 pc in diameter, several SFMC particles in the same clump show nearly identical properties of their environment regardless of their ages. 
    \label{fig:evolution_300pc_rad}}
\end{figure*}

To better illustrate this effect, one of the actively star-forming clumps are enlarged to show the distribution of individual SFMC particles in Figure \ref{fig:clump_radius_age}.  
The top left panel of the figure displays the locations of SFMC particles colored by their ages at 13.3 Myr into the high-resolution evolution, plotted on top of a black-and-white face-on projection of SFR surface density from mock ${\rm H}\alpha$ emission in a 0.2 kpc box.  
Here, a mild gradient in the ages of SFMC particles is noticeable.  
That is, old stars tend to be at the outskirts of this star-forming region.  
The trend is clearly visible in the bottom right panel of Figure \ref{fig:clump_radius_age}, radially-averaged ages of SFMC particles as a function of distance from the center of the star-forming clump.
There is about 1 Myr age difference on average between the SFMC particles that are closest to the center of the star-forming clump and the ones that are 0.1 kpc away from it.
This trend may suggest that the SFMC particles have drifted away from their birthplaces.  
One might also conjecture that the trend could be fueled by the different modes of stellar feedback, ionizing radiation ($0 < T < 6$ Myr) and supernova ($4 < T < 6$ Myr), which disperse the surrounding molecular gas.
Readers should note that this observation is in line with the finding in Paper I that the escape fraction from a SFMC particle increases on average from 0.27\% at its birth to 2.1\% at the end of a SFMC particle lifetime, 6 Myrs.  
For these reasons, we speculate that the large scatter and/or the evolution of physical properties in the  vicinity of SFMC particles further aid the breakdown of the traditional Kennicutt-Schmidt type star formation laws at smaller scales.  
We, however, acknowledge that identifying the exact process of interactions between SFMC particles and molecular gas (e.g. drifting, dispersal) would require much higher numerical accuracy to trace trajectories of individual SFMC particles and to resolve individual HII regions around those particles.

By changing the resolution of the previous experiment in Figure \ref{fig:evolution_4pc_rad}, we can examine how the star-forming regions are sampled with larger apertures ($\gtrsim$ 300 pc).
In Figure \ref{fig:evolution_300pc_rad}, we plot the physical quantities in the spheres of 600 pc diameter centered on active SFMC particles as functions of $T$.  
Due to the larger size of the aperture, averaged properties such as (${\rm H}_2$) densities look less scattered, and more clustered around the average values (red dashed lines).  
Because several SFMC particles tend to congregate in a single star-forming clump that is often much smaller than 600 pc, they often exhibit nearly identical environmental properties regardless of their ages (i.e. data points distributed almost horizontally lined up). 
This then results in little or no evolution of the averaged physical properties in each panel (i.e. a flat red dashed lines).  
Therefore, traditional star formation laws still hold in observations made with apertures that are large enough to average over various evolutionary stages of GMCs.   

\subsection{Comparison to the Run Without Stellar Radiation} \label{sec-IV:5-comp}

\begin{figure*}[t]
\epsscale{1.13}
\plotone{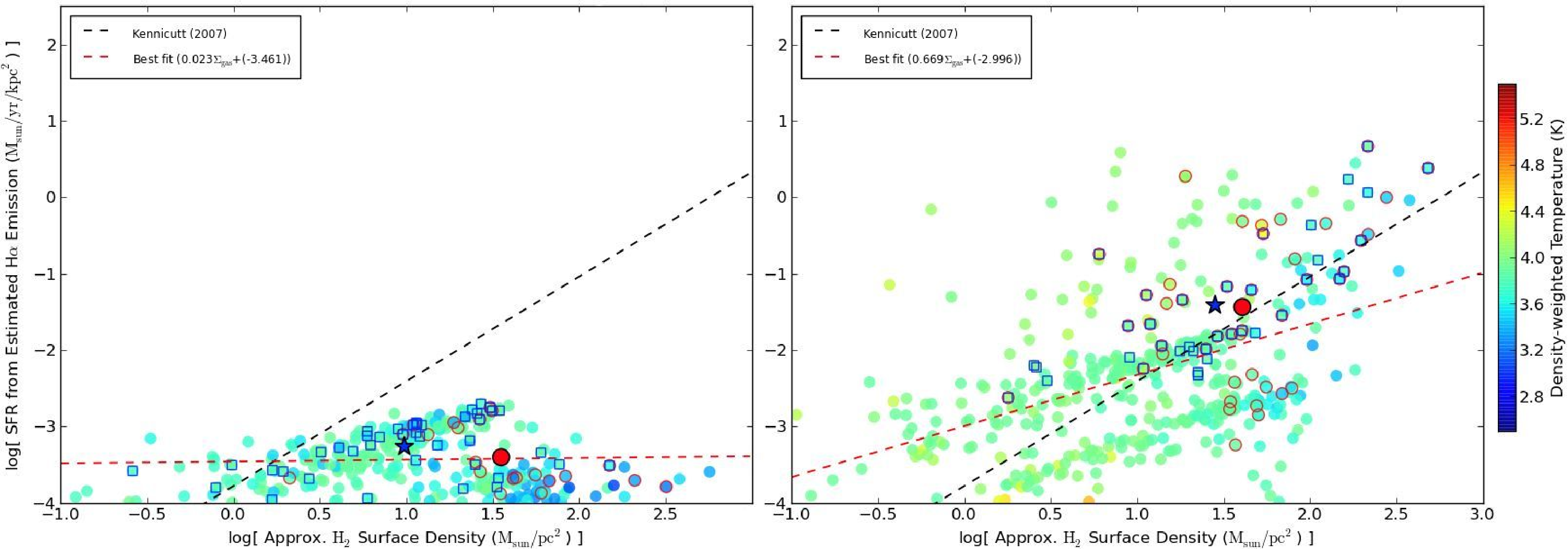}
    \caption{The simulated spatially-resolved Kennicutt-Schmidt plot produced with 75 pc resolution at 13.3 Myr into the high-resolution evolution for the MC-TF ({\it left}, only with supernova feedback) and MC-RTF runs ({\it right}, with both supernova and stellar radiation feedback, same as the lower left panel of Figure \ref{fig:KS_multiple}). See the caption of Figure \ref{fig:KS_multiple} for a detailed description of these plots.  Due to the lack of ionizing radiation in the MC-TF run, the SFR density estimated from ${\rm H}\alpha$ emission is very low in comparison with that of the MC-RTF run.  The color version of this figure is available in the electronic edition.
\label{fig:KS_compare_TF}}
\end{figure*}

So far in \S\ref{sec-IV:5}, we have exclusively focused on the MC-RTF run which includes both stellar radiation and supernova feedback. 
Our study is made possible only via the consistent integration of high spatial resolution, realistic feedback physics and chemistry, and post-production process.
To demonstrate this, we compare our results in \S\ref{sec-IV:5-spatial} to that of the MC-TF run which includes supernova feedback, but not ionizing radiation. 
In Figure \ref{fig:KS_compare_TF}, we have post-processed the MC-TF run in the same manner as the MC-RTF run at the identical timestep (lower left panel of Figure \ref{fig:KS_multiple}). 
Due to the lack of radiation that could have ionized hydrogen in the vicinity of SFMC particles, ${\rm H}\alpha$ emission and SFR density estimated from it are very low in the MC-TF run.
Obviously, supernova feedback alone fails to achieve the reasonable galactic ISM with a sensible density of ionized hydrogen around SFMC particles.
One can thus conclude that post-processing alone is insufficient to produce realistic mock observations like Figure \ref{fig:maps_50pc}, or Kennicutt-Schmidt plots.  
Such maps can be properly generated only by self-consistently including radiation calculation in a high-resolution galaxy formation simulation.  

\section{Summary and Conclusion} \label{sec-IV:6}

Using a comprehensive high-resolution simulation of a dwarf-sized galaxy including a sophisticated model of stellar feedback, we have examined the globally-averaged star formation, and the spatially-resolved star formation relation on a simulated galactic disk.
Our goal has been to investigate what prompts the spatially-resolved star formation relation observed in local galaxies, and why it breaks down at small scales.  
The machinery developed in Paper I is ideal for this purpose since it self-consistently computes the interaction of the ionizing stellar radiation with neighboring gas clouds. 
Our major findings are as follows.  

\begin{enumerate}
\item {\it Self-regulated Galactic Star Formation:}
Our new implementation of stellar feedback includes ionizing radiation as well as supernova explosions, and we handle ionizing radiation by solving the radiative transfer equation rather than by a subgrid model (\S\ref{sec-IV:2}).
Joined with high numerical resolution of 3.8 pc, the realistic description of stellar feedback helps to self-regulate star formation (\S\ref{sec-IV:4-SF}).
Photoheating by stellar radiation retains hot dense gas above the density threshold for SFMC particle creation, $10^3 \,\,{\rm cm}^{-3}$, which otherwise would have been unstable against Jeans fragmentation and deposited into SFMC particles (\S\ref{sec-IV:4-ISM}). 

\item {\it Simulated Observation of Star Formation Relation:}
Because we have self-consistently calculated the location of ionized gas, we are able to make spatially-resolved mock observations of galactic star formation tracers, such as ${\rm H}\alpha$ emission.  
It is also feasible to observe how stellar feedback manifests itself in the correlation between ionized and molecular  gas  (\S\ref{sec-IV:5-obs}).
The reported mock observations are possible only because the following key ingredients are systematically integrated: {\it (a)} sufficient resolution both in space and time, {\it (b)} the realistic stellar feedback model including ionizing radiation, combined with primordial chemistry, and {\it (c)} the post-production procedure to evaluate ${\rm H}\alpha$ emission and ${\rm H}_2$ fraction on a galactic disk.
The realistic stellar feedback model established in this series of papers provides us with an unprecedented insight into the physics of galactic star formation.

\item {\it Spatially-resolved Star Formation Relation:}
Applying our mock observation techniques to the disk in a galactic halo of $2.3 \times 10^{11} M_{\odot}$, we find that the spatially-resolved star formation correlation between SFR density (estimated from mock ${\rm H}\alpha$ emission) and ${\rm H}_2$ density shows large scatter, especially at high resolutions of $\lesssim$ 75 pc.
We also reproduce the phenomenon that correlation of the Kennicutt-Schmidt type observed at $\sim$ kpc scales becomes loose at small scales that are comparable to the size of GMCs, and varies systematically depending on whether apertures are centered on ${\rm H}\alpha$ or ${\rm H}_2$ peaks.
This is because an aperture of GMC size only captures particular stages of GMC evolution, and because ${\rm H}\alpha$ traces hot gas around star-forming regions and is displaced from the ${\rm H}_2$ peaks themselves (\S\ref{sec-IV:5-spatial}).  

\item {\it Evolving Environment Around Star-forming Particles:}
By examining the evolving environment around SFMC particles, we speculate that the large scatter and/or the evolution of physical properties in the very vicinity of SFMC particles are caused by a combination of stars drifting from their birthplaces, and molecular clouds being dispersed via stellar feedback (\S\ref{sec-IV:5-environ}).
These factors may further aid the breakdown of the traditional star formation laws at small scales.  

\end{enumerate} 

Building upon the SFMC scheme developed in this series of papers, we plan to work on more energetic supernova feedback.  
In the reported study, we adopted a relatively small value for supernova feedback energy which is only marginally effective in stopping the unimpeded collapse of star-forming gas clumps.  
By using more energetic supernova energy \cite[e.g.][]{2008ApJ...673..810T}, we will investigate the relative importance of ionized gas pressure and supernova explosions as sources of stellar feedback.  
We also aim to include the effect of young runaway stars \citep[e.g.][]{2010MNRAS.404.2151C} which could potentially alter the distribution of ionized gas around SFMC particles and enhance the escape of ionizing photons.

\vspace{1 mm}

\acknowledgments

J. K. thanks Joel Primack, Chao-chin Yang, and an anonymous referee for providing insightful comments and valuable advice.    
M. R. K. acknowledges support from an Alfred P. Sloan Fellowship, from the NSF through grant CAREER-0955300, and from NASA through a Chandra Space Telescope Grant and through Astrophysics Theory and Fundamental Physics Grant NNX09AK31G.
J. H. W. gratefully acknowledges support from the NSF Grant AST-1211626.  
M. J. T. gratefully acknowledges support from the NSF Grant OCI-1048505. 
N. J. G. is supported by a Graduate Research Fellowship from the NSF.
The examination of the simulation data and the post-production analysis are greatly aided by an AMR analysis toolkit {\it yt} \citep{2011ApJS..192....9T}.
This work used the Extreme Science and Engineering Discovery Environment (XSEDE), which is supported by NSF grant OCI-1053575.
The authors acknowledge the Texas Advanced Computing Center (TACC) at the University of Texas at Austin for providing high-performance computing resources that have contributed to the research results reported within this paper.

\begin{appendix}

\section{A. Comparison of Estimated ${\rm H}\alpha$ Emission and Actual Star Formation Rate: Low Resolution} \label{sec:appendix-A}

\begin{figure*}[t]
\epsscale{1.17}
\plotone{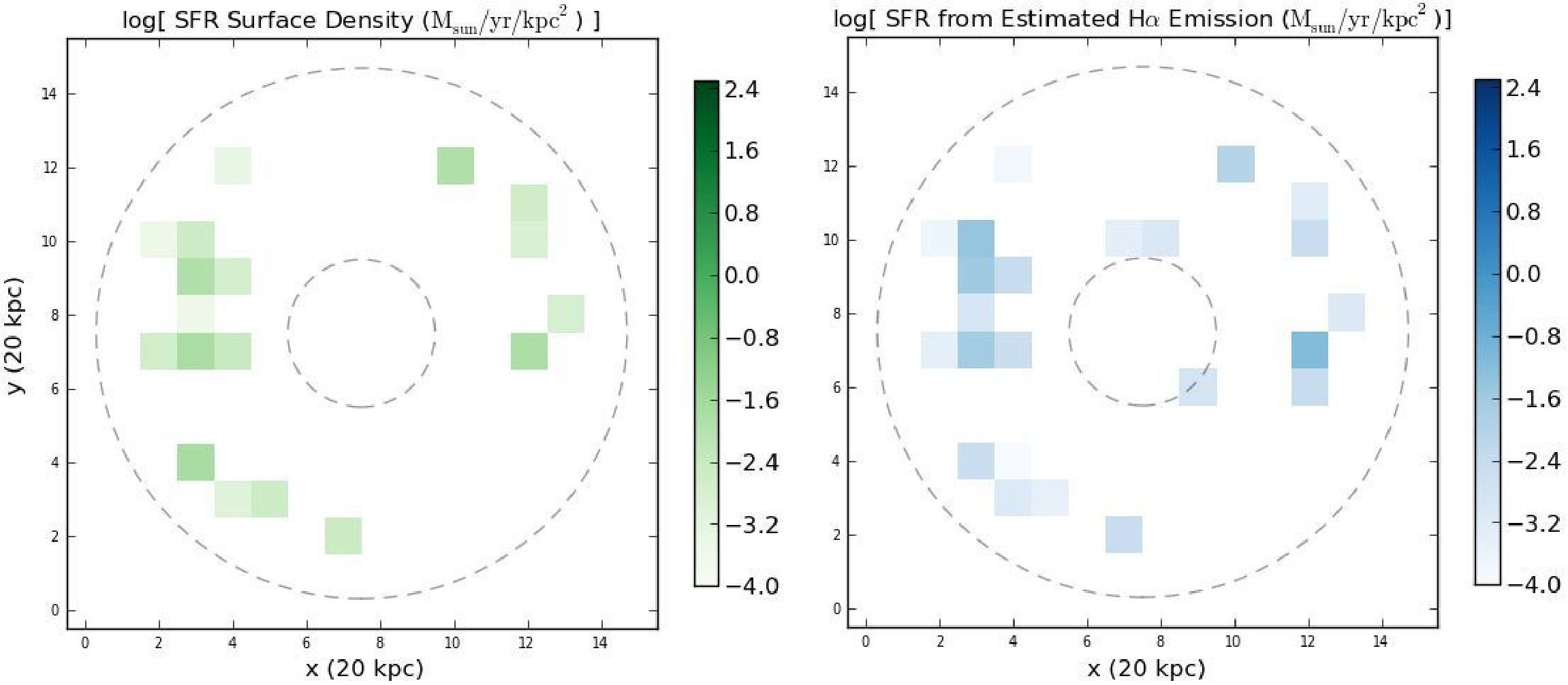}
    \caption{``Actual'' SFR surface density ({\it left}, in $M_{\odot}\,{\rm yr}^{-1}\,{\rm kpc}^{-2}$) and the ``estimated" SFR surface density evaluated from mock ${\rm H}\alpha$ emission ({\it right}).  Shown at 13.3 Myr into the high-resolution evolution of the MC-RTF run in a 20 kpc box centered on the galactic center.  The same mock observation technique as in Figures \ref{fig:maps_preview} and \ref{fig:maps_50pc} is employed with uniform 1.2 kpc resolution.
\label{fig:Halpha_SFR_compare}}
\end{figure*}

In this section we discuss whether the ``estimated'' value of star formation rate (SFR) from the mock ${\rm H}\alpha$ emission is reliable when compared with the ``actual'' value of SFR recorded in the simulation.
Figure \ref{fig:Halpha_SFR_compare} shows the actual SFR surface density, $\Sigma_{\rm SFR}$, and the estimated SFR surface density, $\Sigma_{\rm SFR,\,est}$,  in a 20 kpc box.  
Here the same mock observation technique as in Figures \ref{fig:maps_preview} and \ref{fig:maps_50pc} is employed with much larger 1.2 kpc resolution, large enough to enclose both ${\rm H}\alpha$ and ``actual'' SFR in one pixel.
The actual SFR is found by locating the SFMC particles that are younger than 3 Myrs and calculating their surface density in each of $(1.2\,\, {\rm kpc})^2$ resolution elements on the disk plane. 
By averaging the non-zero resolution elements  of Figure \ref{fig:Halpha_SFR_compare} we find that the ``actual'' global SFR surface density is $\overline \Sigma_{\rm SFR} = 2.63 \times 10^{-3} \,\,M_{\odot}\,{\rm yr}^{-1}\,{\rm kpc}^{-2}$ whereas the ``estimated'' global SFR surface density from ${\rm H}_{\alpha}$ is $\overline \Sigma_{\rm SFR,\,est} = 2.34 \times 10^{-3} \,\,M_{\odot}\,{\rm yr}^{-1}\,{\rm kpc}^{-2}$.  
The difference between the two is only about 11\%, thereby confirming that our ${\rm H}\alpha$ emission estimate, Eq.(\ref{eq:Ha_by_recomb})$+$(\ref{eq:Ha_by_colexc}), and the ${\rm H}\alpha$-to-SFR conversion factor, Eq.(\ref{eq:Ha_conversion}), are reliable.  
In other words, this test verifies that, in simulated observations at low resolution ($> 1 \, {\rm kpc}$), our ${\rm H}\alpha$ estimate closely follows the actual star formation rate recorded in the simulation. 
We note that $\overline \Sigma_{\rm SFR}$ is slightly larger than $\overline \Sigma_{\rm SFR,\,est}$ in our method.Ê It is because $\overline \Sigma_{\rm SFR}$ is averaged over a small number of resolution elements, while $\overline \Sigma_{\rm SFR,\,est}$ is over a larger set of cells that are illuminated by the stellar radiation.ÊÊ

\section{B. On the Use of Krumholz et al. Equilibrium Model To Determine Molecular Gas Content} \label{sec:appendix-B}

In this section we discuss whether the use of \cite{2008ApJ...689..865K, 2009ApJ...693..216K} equilibrium model (hereafter KMT model) is adequate in our experiment to determine molecular gas content.
First, we examine whether it is reasonable to apply the KMT model on scales below 65 pc tested by \cite{2011ApJ...729...36K}. 
More recent theoretical and observational studies suggest that it is.  
Observationally, \cite{2012ApJ...748...75L} report observations of the nearby Perseus Molecular Cloud with resolution of 0.4 pc, and find that the KMT model provides an excellent fit to the data down to these scales (see their Figures 11 and 12).    
Theoretically, \cite{2012ApJ...759....9K} investigates the importance of non-equilibrium effects on the ${\rm H_2}$ fraction, and finds that they are negligible at metallicities above $\sim10\%$ of the solar value, because the equilibration time remains smaller than the free-fall time in this metallicity range. 

Second, we consider whether it is reasonable to apply the KMT model to clouds in the process of being destroyed by a radiation field.  
We follow the ionizing radiation and solve for the ionization chemistry, but do not directly trace the photo-dissociating part of the radiation field from $912 - 1100\,\, \AA$. 
These photons will produce a photo-dissociation front that extends beyond the ionization front, which we are neglecting.
Whether this is reasonable or not depends on the hardness of the spectrum. 
For a young stellar population drawn from a fully-sampled initial mass function (IMF), a number of authors have shown that the dissociation front always remains trapped between the ionization front and the shock front that precedes the expanding HII region \citep{1996ApJ...468..269D, 2007ApJ...671..518K, 2011MNRAS.414.1747A}.  
In this situation, the dissociation region is neither important dynamically, nor in terms of the mass budget, and thus there is no need to consider it.
 
The approximation may become more problematic for old star cluster particles, where the most massive stars have died off but less massive ones remain, softening the spectrum.  
However, this effect is only significant for ages $\gtrsim$ 7-8 Myr, which we do not examine in the reported study since their contribution to ${\rm H}\alpha$ emission is small.  
We nonetheless warn that the {\it collective} effects of many such older star particles are ignored in our experiment.
By adopting the KMT model, we implicitly adopt a single, uniform FUV radiation field throughout the galaxy, rather than accounting for the local enhancement of the FUV in star-forming regions where large $\sim$ 10 Myr old stellar populations are present.  
This effect might reduce the ${\rm H_2}$ fraction compared to our fiducial model, although we note that the effect should be partially countered by an increase in mean density of the atomic cold neutral medium induced by the harder radiation field \citep[see][]{2008ApJ...689..865K, 2009ApJ...693..216K}.  
Treating this problem accurately would require not just a time-dependent treatment of the ${\rm H_2}$  chemistry, but also a time-dependent solution to the radiative transfer equation for FUV photons. 

\end{appendix}


\begin{thebibliography}{63}
\expandafter\ifx\csname natexlab\endcsname\relax\def\natexlab#1{#1}\fi

\bibitem[{{Abel} \& {Wandelt}(2002)}]{2002MNRAS.330L..53A}
{Abel}, T., \& {Wandelt}, B.~D. 2002, \mnras, 330, L53

\bibitem[{{Aggarwal}(1983)}]{1983MNRAS.202P..15A}
{Aggarwal}, K.~M. 1983, \mnras, 202, 15P

\bibitem[{{Arthur} {et~al.}(2011){Arthur}, {Henney}, {Mellema}, {de Colle}, \&
  {V{\'a}zquez-Semadeni}}]{2011MNRAS.414.1747A}
{Arthur}, S.~J., {Henney}, W.~J., {Mellema}, G., {de Colle}, F., \&
  {V{\'a}zquez-Semadeni}, E. 2011, \mnras, 414, 1747

\bibitem[{{Bigiel} {et~al.}(2010){Bigiel}, {Leroy}, {Walter}, {Blitz},
  {Brinks}, {de Blok}, \& {Madore}}]{2010AJ....140.1194B}
{Bigiel}, F., {Leroy}, A., {Walter}, F., {Blitz}, L., {Brinks}, E., {de Blok},
  W.~J.~G., \& {Madore}, B. 2010, \aj, 140, 1194

\bibitem[{{Bigiel} {et~al.}(2008){Bigiel}, {Leroy}, {Walter}, {Brinks}, {de
  Blok}, {Madore}, \& {Thornley}}]{2008AJ....136.2846B}
{Bigiel}, F., {Leroy}, A., {Walter}, F., {Brinks}, E., {de Blok}, W.~J.~G.,
  {Madore}, B., \& {Thornley}, M.~D. 2008, \aj, 136, 2846

\bibitem[{{Bigiel} {et~al.}(2011){Bigiel}, {Leroy}, {Walter}, {Brinks}, {de
  Blok}, {Kramer}, {Rix}, {Schruba}, {Schuster}, {Usero}, \&
  {Wiesemeyer}}]{2011ApJ...730L..13B}
{Bigiel}, F., {Leroy}, A.~K., {Walter}, F., {Brinks}, E., {de Blok}, W.~J.~G.,
  {Kramer}, C., {Rix}, H.~W., {Schruba}, A., {Schuster}, K.-F., {Usero}, A., \&
  {Wiesemeyer}, H.~W. 2011, \apjl, 730, L13

\bibitem[{{Bolatto} {et~al.}(2011){Bolatto}, {Leroy}, {Jameson}, {Ostriker},
  {Gordon}, {Lawton}, {Stanimirovi{\'c}}, {Israel}, {Madden}, {Hony},
  {Sandstrom}, {Bot}, {Rubio}, {Winkler}, {Roman-Duval}, {van Loon},
  {Oliveira}, \& {Indebetouw}}]{2011ApJ...741...12B}
{Bolatto}, A.~D., {Leroy}, A.~K., {Jameson}, K., {Ostriker}, E., {Gordon}, K.,
  {Lawton}, B., {Stanimirovi{\'c}}, S., {Israel}, F.~P., {Madden}, S.~C.,
  {Hony}, S., {Sandstrom}, K.~M., {Bot}, C., {Rubio}, M., {Winkler}, P.~F.,
  {Roman-Duval}, J., {van Loon}, J.~T., {Oliveira}, J.~M., \& {Indebetouw}, R.
  2011, \apj, 741, 12

\bibitem[{{Bolatto} {et~al.}(2013){Bolatto}, {Wolfire}, \&
  {Leroy}}]{2013arXiv1301.3498B}
{Bolatto}, A.~D., {Wolfire}, M., \& {Leroy}, A.~K. 2013, arXiv:1301.3498

\bibitem[{{Bouch{\'e}} {et~al.}(2007){Bouch{\'e}}, {Cresci}, {Davies},
  {Eisenhauer}, {F{\"o}rster Schreiber}, {Genzel}, {Gillessen}, {Lehnert},
  {Lutz}, {Nesvadba}, {Shapiro}, {Sternberg}, {Tacconi}, {Verma}, {Cimatti},
  {Daddi}, {Renzini}, {Erb}, {Shapley}, \& {Steidel}}]{2007ApJ...671..303B}
{Bouch{\'e}}, N., {Cresci}, G., {Davies}, R., {Eisenhauer}, F., {F{\"o}rster
  Schreiber}, N.~M., {Genzel}, R., {Gillessen}, S., {Lehnert}, M., {Lutz}, D.,
  {Nesvadba}, N., {Shapiro}, K.~L., {Sternberg}, A., {Tacconi}, L.~J., {Verma},
  A., {Cimatti}, A., {Daddi}, E., {Renzini}, A., {Erb}, D.~K., {Shapley}, A.,
  \& {Steidel}, C.~C. 2007, \apj, 671, 303

\bibitem[{{Bournaud} {et~al.}(2010){Bournaud}, {Elmegreen}, {Teyssier},
  {Block}, \& {Puerari}}]{2010MNRAS.409.1088B}
{Bournaud}, F., {Elmegreen}, B.~G., {Teyssier}, R., {Block}, D.~L., \&
  {Puerari}, I. 2010, \mnras, 409, 1088

\bibitem[{{Bryan} \& {Norman}(1997)}]{1997ASPC..123..363B}
{Bryan}, G.~L., \& {Norman}, M.~L. 1997, in Astronomical Society of the Pacific
  Conference Series, Vol. 123, Computational Astrophysics; 12th Kingston
  Meeting on Theoretical Astrophysics, ed. D.~A. {Clarke} \& M.~J. {West},
  363--368

\bibitem[{{Calzetti} {et~al.}(2007){Calzetti}, {Kennicutt}, {Engelbracht},
  {Leitherer}, {Draine}, {Kewley}, {Moustakas}, {Sosey}, {Dale}, {Gordon},
  {Helou}, {Hollenbach}, {Armus}, {Bendo}, {Bot}, {Buckalew}, {Jarrett}, {Li},
  {Meyer}, {Murphy}, {Prescott}, {Regan}, {Rieke}, {Roussel}, {Sheth}, {Smith},
  {Thornley}, \& {Walter}}]{2007ApJ...666..870C}
{Calzetti}, D., {Kennicutt}, R.~C., {Engelbracht}, C.~W., {Leitherer}, C.,
  {Draine}, B.~T., {Kewley}, L., {Moustakas}, J., {Sosey}, M., {Dale}, D.~A.,
  {Gordon}, K.~D., {Helou}, G.~X., {Hollenbach}, D.~J., {Armus}, L., {Bendo},
  G., {Bot}, C., {Buckalew}, B., {Jarrett}, T., {Li}, A., {Meyer}, M.,
  {Murphy}, E.~J., {Prescott}, M., {Regan}, M.~W., {Rieke}, G.~H., {Roussel},
  H., {Sheth}, K., {Smith}, J.~D.~T., {Thornley}, M.~D., \& {Walter}, F. 2007,
  \apj, 666, 870

\bibitem[{{Ceverino} {et~al.}(2010){Ceverino}, {Dekel}, \&
  {Bournaud}}]{2010MNRAS.404.2151C}
{Ceverino}, D., {Dekel}, A., \& {Bournaud}, F. 2010, \mnras, 404, 2151

\bibitem[{{Dong} \& {Draine}(2011)}]{2011ApJ...727...35D}
{Dong}, R., \& {Draine}, B.~T. 2011, \apj, 727, 35

\bibitem[{{Draine} \& {Bertoldi}(1996)}]{1996ApJ...468..269D}
{Draine}, B.~T., \& {Bertoldi}, F. 1996, \apj, 468, 269

\bibitem[{{Feldmann} {et~al.}(2011){Feldmann}, {Gnedin}, \&
  {Kravtsov}}]{2011ApJ...732..115F}
{Feldmann}, R., {Gnedin}, N.~Y., \& {Kravtsov}, A.~V. 2011, \apj, 732, 115

\bibitem[{{Feldmann} {et~al.}(2012{\natexlab{a}}){Feldmann}, {Gnedin}, \&
  {Kravtsov}}]{2012ApJ...747..124F}
---. 2012{\natexlab{a}}, \apj, 747, 124

\bibitem[{{Feldmann} {et~al.}(2012{\natexlab{b}}){Feldmann}, {Gnedin}, \&
  {Kravtsov}}]{2012ApJ...758..127F}
---. 2012{\natexlab{b}}, \apj, 758, 127

\bibitem[{{Hopkins} {et~al.}(2011){Hopkins}, {Quataert}, \&
  {Murray}}]{2011MNRAS.417..950H}
{Hopkins}, P.~F., {Quataert}, E., \& {Murray}, N. 2011, \mnras, 417, 950

\bibitem[{{Kennicutt}(1989)}]{1989ApJ...344..685K}
{Kennicutt}, Jr., R.~C. 1989, \apj, 344, 685

\bibitem[{{Kennicutt}(1998)}]{1998ApJ...498..541K}
---. 1998, \apj, 498, 541

\bibitem[{{Kennicutt} {et~al.}(2007){Kennicutt}, {Calzetti}, {Walter}, {Helou},
  {Hollenbach}, {Armus}, {Bendo}, {Dale}, {Draine}, {Engelbracht}, {Gordon},
  {Prescott}, {Regan}, {Thornley}, {Bot}, {Brinks}, {de Blok}, {de Mello},
  {Meyer}, {Moustakas}, {Murphy}, {Sheth}, \& {Smith}}]{2007ApJ...671..333K}
{Kennicutt}, Jr., R.~C., {Calzetti}, D., {Walter}, F., {Helou}, G.,
  {Hollenbach}, D.~J., {Armus}, L., {Bendo}, G., {Dale}, D.~A., {Draine},
  B.~T., {Engelbracht}, C.~W., {Gordon}, K.~D., {Prescott}, M.~K.~M., {Regan},
  M.~W., {Thornley}, M.~D., {Bot}, C., {Brinks}, E., {de Blok}, E., {de Mello},
  D., {Meyer}, M., {Moustakas}, J., {Murphy}, E.~J., {Sheth}, K., \& {Smith},
  J.~D.~T. 2007, \apj, 671, 333

\bibitem[{{Kim} {et~al.}(2013){Kim}, {Krumholz}, {Wise}, {Turk}, {Goldbaum}, \&
  {Abel}}]{2013ApJ...775..109K}
{Kim}, J.-H., {Krumholz}, M.~R., {Wise}, J.~H., {Turk}, M.~J., {Goldbaum},
  N.~J., \& {Abel}, T. 2013, \apj, 775, 109

\bibitem[{{Kim} {et~al.}(2009){Kim}, {Wise}, \& {Abel}}]{2009ApJ...694L.123K}
{Kim}, J.-H., {Wise}, J.~H., \& {Abel}, T. 2009, \apjl, 694, L123

\bibitem[{{Kim} {et~al.}(2011){Kim}, {Wise}, {Alvarez}, \&
  {Abel}}]{2011ApJ...738...54K}
{Kim}, J.-H., {Wise}, J.~H., {Alvarez}, M.~A., \& {Abel}, T. 2011, \apj, 738,
  54

\bibitem[{{Komugi} {et~al.}(2012){Komugi}, {Tateuchi}, {Motohara}, {Takagi},
  {Iono}, {Kaneko}, {Ueda}, {Saitoh}, {Kato}, {Konishi}, {Koshida}, {Morokuma},
  {Takahashi}, {Tanab{\'e}}, \& {Yoshii}}]{2012ApJ...757..138K}
{Komugi}, S., {Tateuchi}, K., {Motohara}, K., {Takagi}, T., {Iono}, D.,
  {Kaneko}, H., {Ueda}, J., {Saitoh}, T.~R., {Kato}, N., {Konishi}, M.,
  {Koshida}, S., {Morokuma}, T., {Takahashi}, H., {Tanab{\'e}}, T., \&
  {Yoshii}, Y. 2012, \apj, 757, 138

\bibitem[{{Koyama} \& {Inutsuka}(2002)}]{2002ApJ...564L..97K}
{Koyama}, H., \& {Inutsuka}, S. 2002, \apjl, 564, L97

\bibitem[{{Krumholz}(2012)}]{2012ApJ...759....9K}
{Krumholz}, M.~R. 2012, \apj, 759, 9

\bibitem[{{Krumholz} {et~al.}(2012){Krumholz}, {Dekel}, \&
  {McKee}}]{2012ApJ...745...69K}
{Krumholz}, M.~R., {Dekel}, A., \& {McKee}, C.~F. 2012, \apj, 745, 69

\bibitem[{{Krumholz} \& {Gnedin}(2011)}]{2011ApJ...729...36K}
{Krumholz}, M.~R., \& {Gnedin}, N.~Y. 2011, \apj, 729, 36

\bibitem[{{Krumholz} {et~al.}(2008){Krumholz}, {McKee}, \&
  {Tumlinson}}]{2008ApJ...689..865K}
{Krumholz}, M.~R., {McKee}, C.~F., \& {Tumlinson}, J. 2008, \apj, 689, 865

\bibitem[{{Krumholz} {et~al.}(2009{\natexlab{a}}){Krumholz}, {McKee}, \&
  {Tumlinson}}]{2009ApJ...693..216K}
---. 2009{\natexlab{a}}, \apj, 693, 216

\bibitem[{{Krumholz} {et~al.}(2009{\natexlab{b}}){Krumholz}, {McKee}, \&
  {Tumlinson}}]{2009ApJ...699..850K}
---. 2009{\natexlab{b}}, \apj, 699, 850

\bibitem[{{Krumholz} {et~al.}(2007){Krumholz}, {Stone}, \&
  {Gardiner}}]{2007ApJ...671..518K}
{Krumholz}, M.~R., {Stone}, J.~M., \& {Gardiner}, T.~A. 2007, \apj, 671, 518

\bibitem[{{Krumholz} \& {Thompson}(2007)}]{2007ApJ...669..289K}
{Krumholz}, M.~R., \& {Thompson}, T.~A. 2007, \apj, 669, 289

\bibitem[{{Kuhlen} {et~al.}(2012){Kuhlen}, {Krumholz}, {Madau}, {Smith}, \&
  {Wise}}]{2012ApJ...749...36K}
{Kuhlen}, M., {Krumholz}, M.~R., {Madau}, P., {Smith}, B.~D., \& {Wise}, J.
  2012, \apj, 749, 36

\bibitem[{{Lee} {et~al.}(2012){Lee}, {Stanimirovi{\'c}}, {Douglas}, {Knee}, {Di
  Francesco}, {Gibson}, {Begum}, {Grcevich}, {Heiles}, {Korpela}, {Leroy},
  {Peek}, {Pingel}, {Putman}, \& {Saul}}]{2012ApJ...748...75L}
{Lee}, M.-Y., {Stanimirovi{\'c}}, S., {Douglas}, K.~A., {Knee}, L.~B.~G., {Di
  Francesco}, J., {Gibson}, S.~J., {Begum}, A., {Grcevich}, J., {Heiles}, C.,
  {Korpela}, E.~J., {Leroy}, A.~K., {Peek}, J.~E.~G., {Pingel}, N.~M.,
  {Putman}, M.~E., \& {Saul}, D. 2012, \apj, 748, 75

\bibitem[{{Leroy} {et~al.}(2013){Leroy}, {Walter}, {Sandstrom}, {Schruba},
  {Munoz-Mateos}, {Bigiel}, {Bolatto}, {Brinks}, {de Blok}, {Meidt}, {Rix},
  {Rosolowsky}, {Schinnerer}, {Schuster}, \& {Usero}}]{2013arXiv1301.2328L}
{Leroy}, A.~K., {Walter}, F., {Sandstrom}, K., {Schruba}, A., {Munoz-Mateos},
  J.-C., {Bigiel}, F., {Bolatto}, A., {Brinks}, E., {de Blok}, W.~J.~G.,
  {Meidt}, S., {Rix}, H.-W., {Rosolowsky}, E., {Schinnerer}, E., {Schuster},
  K.-F., \& {Usero}, A. 2013, arXiv:1301.2328

\bibitem[{{Liu} {et~al.}(2011){Liu}, {Koda}, {Calzetti}, {Fukuhara}, \&
  {Momose}}]{2011ApJ...735...63L}
{Liu}, G., {Koda}, J., {Calzetti}, D., {Fukuhara}, M., \& {Momose}, R. 2011,
  \apj, 735, 63

\bibitem[{{Luridiana} {et~al.}(2003){Luridiana}, {Peimbert}, {Peimbert}, \&
  {Cervi{\~n}o}}]{2003ApJ...592..846L}
{Luridiana}, V., {Peimbert}, A., {Peimbert}, M., \& {Cervi{\~n}o}, M. 2003,
  \apj, 592, 846

\bibitem[{{Matzner}(2002)}]{2002ApJ...566..302M}
{Matzner}, C.~D. 2002, \apj, 566, 302

\bibitem[{{McKee} \& {Krumholz}(2010)}]{2010ApJ...709..308M}
{McKee}, C.~F., \& {Krumholz}, M.~R. 2010, \apj, 709, 308

\bibitem[{{Murray} \& {Rahman}(2010)}]{2010ApJ...709..424M}
{Murray}, N., \& {Rahman}, M. 2010, \apj, 709, 424

\bibitem[{{Narayanan} {et~al.}(2011){Narayanan}, {Krumholz}, {Ostriker}, \&
  {Hernquist}}]{2011MNRAS.418..664N}
{Narayanan}, D., {Krumholz}, M., {Ostriker}, E.~C., \& {Hernquist}, L. 2011,
  \mnras, 418, 664

\bibitem[{{Narayanan} {et~al.}(2012){Narayanan}, {Krumholz}, {Ostriker}, \&
  {Hernquist}}]{2012MNRAS.421.3127N}
{Narayanan}, D., {Krumholz}, M.~R., {Ostriker}, E.~C., \& {Hernquist}, L. 2012,
  \mnras, 421, 3127

\bibitem[{{Norman} {et~al.}(2007){Norman}, {Bryan}, {Harkness}, {Bordner},
  {Reynolds}, {O'Shea}, \& {Wagner}}]{2007arXiv0705.1556N}
{Norman}, M.~L., {Bryan}, G.~L., {Harkness}, R., {Bordner}, J., {Reynolds}, D.,
  {O'Shea}, B., \& {Wagner}, R. 2007, ArXiv e-prints, 705, arXiv:0705.1556

\bibitem[{{Onodera} {et~al.}(2010){Onodera}, {Kuno}, {Tosaki}, {Kohno},
  {Nakanishi}, {Sawada}, {Muraoka}, {Komugi}, {Miura}, {Kaneko}, {Hirota}, \&
  {Kawabe}}]{2010ApJ...722L.127O}
{Onodera}, S., {Kuno}, N., {Tosaki}, T., {Kohno}, K., {Nakanishi}, K.,
  {Sawada}, T., {Muraoka}, K., {Komugi}, S., {Miura}, R., {Kaneko}, H.,
  {Hirota}, A., \& {Kawabe}, R. 2010, \apjl, 722, L127

\bibitem[{{Robertson} \& {Kravtsov}(2008)}]{2008ApJ...680.1083R}
{Robertson}, B.~E., \& {Kravtsov}, A.~V. 2008, \apj, 680, 1083

\bibitem[{{Schmidt}(1959)}]{1959ApJ...129..243S}
{Schmidt}, M. 1959, \apj, 129, 243

\bibitem[{{Schruba} {et~al.}(2011){Schruba}, {Leroy}, {Walter}, {Bigiel},
  {Brinks}, {de Blok}, {Dumas}, {Kramer}, {Rosolowsky}, {Sandstrom},
  {Schuster}, {Usero}, {Weiss}, \& {Wiesemeyer}}]{2011AJ....142...37S}
{Schruba}, A., {Leroy}, A.~K., {Walter}, F., {Bigiel}, F., {Brinks}, E., {de
  Blok}, W.~J.~G., {Dumas}, G., {Kramer}, C., {Rosolowsky}, E., {Sandstrom},
  K., {Schuster}, K., {Usero}, A., {Weiss}, A., \& {Wiesemeyer}, H. 2011, \aj,
  142, 37

\bibitem[{{Schruba} {et~al.}(2010){Schruba}, {Leroy}, {Walter}, {Sandstrom}, \&
  {Rosolowsky}}]{2010ApJ...722.1699S}
{Schruba}, A., {Leroy}, A.~K., {Walter}, F., {Sandstrom}, K., \& {Rosolowsky},
  E. 2010, \apj, 722, 1699

\bibitem[{{Stinson} {et~al.}(2012){Stinson}, {Brook}, {Macci{\`o}}, {Wadsley},
  {Quinn}, \& {Couchman}}]{2012arXiv1208.0002S}
{Stinson}, G., {Brook}, C., {Macci{\`o}}, A.~V., {Wadsley}, J., {Quinn}, T.~R.,
  \& {Couchman}, H.~M.~P. 2012, ArXiv e-prints, 1208, arXiv:1208.0002

\bibitem[{{Stone} \& {Norman}(1992{\natexlab{a}})}]{1992ApJS...80..753S}
{Stone}, J.~M., \& {Norman}, M.~L. 1992{\natexlab{a}}, \apjs, 80, 753

\bibitem[{{Stone} \& {Norman}(1992{\natexlab{b}})}]{1992ApJS...80..791S}
---. 1992{\natexlab{b}}, \apjs, 80, 791

\bibitem[{{Sutherland} \& {Dopita}(1993)}]{1993ApJS...88..253S}
{Sutherland}, R.~S., \& {Dopita}, M.~A. 1993, \apjs, 88, 253

\bibitem[{{Tasker} \& {Bryan}(2008)}]{2008ApJ...673..810T}
{Tasker}, E.~J., \& {Bryan}, G.~L. 2008, \apj, 673, 810

\bibitem[{{Teyssier} {et~al.}(2010){Teyssier}, {Chapon}, \&
  {Bournaud}}]{2010ApJ...720L.149T}
{Teyssier}, R., {Chapon}, D., \& {Bournaud}, F. 2010, \apjl, 720, L149

\bibitem[{{Turk} {et~al.}(2011){Turk}, {Smith}, {Oishi}, {Skory}, {Skillman},
  {Abel}, \& {Norman}}]{2011ApJS..192....9T}
{Turk}, M.~J., {Smith}, B.~D., {Oishi}, J.~S., {Skory}, S., {Skillman}, S.~W.,
  {Abel}, T., \& {Norman}, M.~L. 2011, \apjs, 192, 9

\bibitem[{{Whalen} \& {Norman}(2006)}]{2006ApJS..162..281W}
{Whalen}, D., \& {Norman}, M.~L. 2006, \apjs, 162, 281

\bibitem[{{Whitworth}(1979)}]{1979MNRAS.186...59W}
{Whitworth}, A. 1979, \mnras, 186, 59

\bibitem[{{Wise} \& {Abel}(2011)}]{2011MNRAS.414.3458W}
{Wise}, J.~H., \& {Abel}, T. 2011, \mnras, 414, 3458

\bibitem[{{Wolfire} {et~al.}(2010){Wolfire}, {Hollenbach}, \&
  {McKee}}]{2010ApJ...716.1191W}
{Wolfire}, M.~G., {Hollenbach}, D., \& {McKee}, C.~F. 2010, \apj, 716, 1191

\bibitem[{{Wong} \& {Blitz}(2002)}]{2002ApJ...569..157W}
{Wong}, T., \& {Blitz}, L. 2002, \apj, 569, 157

\end{thebibliography}
\end{document}